\documentclass{SCIS2018}
\usepackage{microtype}%if unwanted, comment out or use option "draft"

\usepackage{latexsym}
\usepackage{proof}

\usepackage{amsfonts}
\usepackage{graphicx}
\usepackage{booktabs}
\usepackage{wrapfig,lipsum,booktabs}
\usepackage{semantic}
\usepackage{isabelle,isabellesym}

\usepackage{adjustbox}

\newcommand{\specrg}[1]{{\color{blue}{#1}}}
\definecolor{proofcolor}{rgb}{0,0.45,0}
\newcommand{\specb}[1]{{\color{proofcolor}${#1}$}}

\newcommand{\sectprefix}{{Section}}

\newcommand{\figprefix}{{Fig.}}

\newcommand{\tableprefix}{{Table}}

\newcommand{\lemmaprefix}{Lemma}
\newcommand{\theoremprefix}{Theorem}
\newcommand{\defprefix}{Definition}

\usepackage{array}
\newcommand{\PreserveBackslash}[1]{\let\temp=\\#1\let\\=\temp}
\newcolumntype{C}[1]{>{\PreserveBackslash\centering}p{#1}}
\newcolumntype{R}[1]{>{\PreserveBackslash\raggedleft}p{#1}}
\newcolumntype{L}[1]{>{\PreserveBackslash\raggedright}p{#1}}

\newcommand{\zipfigaft}{}%{\vspace{-6pt}}
\newcommand{\zipssect}{}%{\vspace{-6pt}}
\isabellestyle{it}

\let\endisabellecode=\endisabellecode

\newcommand{\llbrace}{\lbrace\mkern-4.5mu\mid}
\newcommand{\rrbrace}{\mid\mkern-4.5mu\rbrace}

\newcommand{\slang}{$\pi$-Core} %{$\texttt{SPECURITY}$}

\newcommand{\cmdfont}[1]{\textbf{#1}} %{\mathsf{#1}}

\newcommand{\cmdbasic}[1]{\cmdfont{Basic} \ {#1}}
\newcommand{\cmdseq}[2]{{#1};;{#2}}
\newcommand{\cmdcond}[3]{\cmdfont{Cond} \ {#1} \ {#2} \ {#3}}
\newcommand{\cmdwhile}[2]{\cmdfont{While} \ {#1} \ {#2}}
\newcommand{\cmdawait}[2]{\cmdfont{Await} \ {#1} \ {#2}}
\newcommand{\cmdnondt}[1]{\cmdfont{Nondt} \ {#1}}
\newcommand{\cmdnone}{\perp}

\newcommand{\event}[1]{\cmdfont{Event} \ {#1}}
\newcommand{\anonevt}[1]{\lfloor{#1}\rfloor}

\newcommand{\evtsystwo}[2]{\{{#1}, \ ... ,\ {#2}\}}
\newcommand{\evtsysdef}{\evtsystwo{\symbEvt_0}{\symbEvt_n}}
\newcommand{\evtseq}[2]{{#1}\oplus{#2}}

\newcommand{\parsysc}{\symbCore \rightarrow \symbevtsys}
\newcommand{\stmtirq}[2]{{#1} $\blacktriangleright$ {#2}}
\newcommand{\stmtirqa}[2]{{#1} \blacktriangleright {#2}}
\newcommand{\stmtatom}[1]{\textbf{ATOM} \ {#1} \ \textbf{END}}
\newcommand{\stmtawait}[2]{\textbf{AWAIT} \ {#1} \ \textbf{THEN} \ {#2} \ \textbf{END}}

\newcommand{\stmtevent}[3]{\textbf{EVENT} \ {#1} \ \textbf{WHEN} \ {#2} \ \textbf{THEN} \ {#3} \ \textbf{END}}

\newcommand{\symbprog}{P}
\newcommand{\symbbexp}{b}
\newcommand{\symbEvt}{\mathcal{E}}
\newcommand{\symbevt}{ev}
\newcommand{\symbevtbd}{\alpha}
\newcommand{\symbevtsys}{\mathcal{S}}
\newcommand{\symbpes}{\mathcal{PS}}

\newcommand{\symbstate}{s}

\newcommand{\symbevtctx}{x}

\newcommand{\symbconf}{\mathcal{C}}
\newcommand{\symbpcomp}{c}

\newcommand{\symbact}{t}
\newcommand{\symbCore}{\mathcal{K}}
\newcommand{\symbcore}{\kappa}
\newcommand{\symbactk}{\delta}

\newcommand{\actk}[2]{{#1}@{#2}}

\newcommand{\symbspec}{\sharp}
\newcommand{\tran}[1]{\stackrel{{#1}}{\longrightarrow}}   
\newcommand{\trank}[2]{\stackrel{\actk{#1}{#2}}{\longrightarrow}}

\newcommand{\myinfer}[2]{
\begin{tabular}{l}
  \textsc{[#1]} \\
  {#2}
\end{tabular}}

\newcommand{\evtran}{\stackrel{e}{\longrightarrow}}
\newcommand{\compfun}{\Psi}
\newcommand{\symbcomp}{\varpi}
\newcommand{\symbcompk}{\widehat{\symbcomp}}
\newcommand{\compsim}[2]{{#1} \asymp {#2}}
\newcommand{\compconjoin}[2]{{#1} \propto {#2}}
\newcommand{\serialize}{\lll}
\newcommand{\Serialize}[2]{{#1} \serialize {#2}}
\newcommand{{\compstps}}{\Psi_\symbpes}
\newcommand{{\compstes}}{\Psi_\symbevtsys}
\newcommand{{\compste}}{\Psi_\symbEvt}
\newcommand{{\compstp}}{\Psi_\symbprog}
\newcommand{\assumefun}{A}
\newcommand{\commitfun}{C}
\newcommand{\rgcond}[4]{\langle #1, #2, #3, #4 \rangle}
\newcommand{\rgconddefault}{\rgcond{pre}{R}{G}{pst}}
\newcommand{\RGSAT}[2]{\models {#1} \ \mathbf{sat} \ {#2}}
\newcommand{\rgsat}[2]{\vdash {#1} \ \mathbf{sat} \ {#2}}
\newcommand{\stset}[1]{\llbrace {#1} \rrbrace}
\newcommand{\bfs}[1]{{^o}#1}
\newcommand{\afs}[1]{{^a}#1}
\newcommand{\thes}[1]{\acute{}{#1}}
\makeatletter
\newcommand{\superimpose}[2]{%
  {\ooalign{$#1\@firstoftwo#2$\cr\hfil$#1\@secondoftwo#2$\hfil\cr}}}
\makeatother

\begin{document}

%%%%%%%%%%%%%%%%%%%%%%%%%%%%%%%%%%%%%%%%%%%%%%%%%%%%%%%
%%% Authors do not modify the information below
%%% 作者不需要修改此处信息
\ArticleType{RESEARCH PAPER}
%\SpecialTopic{}
\Year{2018}
\Month{}
\Vol{61}
\No{}
\DOI{}
\ArtNo{}
\ReceiveDate{}
\ReviseDate{}
\AcceptDate{}
\OnlineDate{}
%%%%%%%%%%%%%%%%%%%%%%%%%%%%%%%%%%%%%%%%%%%%%%%%%%%%%%%

%%% title: 标题
%%%   \title{title}{title for citation}
\title{An Event-based Compositional Reasoning Approach for Concurrent Reactive Systems}
{An Event-based Compositional Reasoning Approach for Concurrent Reactive Systems}

%%% Corresponding author: 通信作者
%%%   \author[number]{Full name}{{email@xxx.com}}
%%% General author: 一般作者
%%%   \author[number]{Full name}{}
\author[1,2]{Yongwang Zhao}{{zhaoyw@buaa.edu.cn}}
\author[3]{David San\'{a}n}{}
\author[3]{Fuyuan Zhang}{}
\author[3]{Yang Liu}{}

%%% Author information for page head. 页眉中的作者信息
\AuthorMark{Yongwang Zhao}

%%% Authors for citation. 首页引用中的作者信息
\AuthorCitation{Yongwang Zhao, David San\'{a}n, et al}

%%% Authors' contribution. 同等贡献
%\contributions{Authors A and B have the same contribution to this work.}

%%% Address. 地址
%%%   \address[number]{Affiliation, City {\rm Postcode}, Country}
\address[1]{School of Computer Science and Engineering, Beihang University, Beijing {\rm 100083}, China}
\address[2]{Beijing Advanced Innovation Center for Big Data and Brain Computing, Beihang University, Beijing {\rm 100083}, China}
\address[3]{School of Computer Science and Engineering, Nanyang Technological University, Singapore {\rm 639798}, Singapore}

%%% Abstract. 摘要
\abstract{
Reactive systems are composed of a well defined set of input events that the system reacts with by executing an associated handler to each event. In concurrent environments, event handlers can interact with the execution of other programs such as hardware interruptions in preemptive systems, or other instances of the reactive system in multicore architectures. 
State of the art rely-guarantee based verification frameworks only focus on imperative programs, being difficult to capture in the rely and guarantee relations interactions with possible infinite sequences of event handlers, and the input arguments to event handlers.
In this paper, we propose the formalisation in Isabelle/HOL of an event-based rely-guarantee approach for concurrent reactive systems. 
We develop a {\slang} language which incorporates a concurrent imperative and system specification language by ``events'', and we build a rely-guarantee proof system for {\slang} and prove its soundness. Our approach can deal with multicore and interruptible concurrency. We use two case studies to show this: an interruptible controller for stepper motors and an ARINC 653 multicore kernel, and prove the functional correctness and preservation of invariants of them in Isabelle/HOL. 
}

%%% Keywords. 关键词

\maketitle

%%%%%%%%%%%%%%%%%%%%%%%%%%%%%%%%%%%%%%%%%%%%%%%%%%%%%%%
%%% The main text. 正文部分
%%%%%%%%%%%%%%%%%%%%%%%%%%%%%%%%%%%%%%%%%%%%%%%%%%%%%%%

\section{Introduction}
\label{sect:intro}

Nowadays high-assurance systems are often designed as \textit{concurrent reactive systems} \cite{Aceto07}. One of their key roles is how they behave in their computing environment, i.e., the sequence of commands the system executes under an input event. We call this behaviour \textit{reaction services}. 
Examples of such systems are operating systems (OS), control systems, and communication systems. In this kind of system, how and when environment interactions happen are key aspects of their specification.
%Formal specification and verification at system level are critical for correctness of concurrent systems. 

The rely-guarantee technique~\cite{Jones83,Colle94,Xu97} represents a fundamental approach to compositional reasoning of \textit{concurrent programs} with shared variables. However, concurrent languages used in existing rely-guarantee methods (e.g. \cite{Xu97,Nieto03,LiangFF12}) do not provide a straightforward way to specify and verify the temporal aspect of reactive systems: the how and when. For instance, if we consider calls to the services offered by an operating system as input events, we can provide the specification handler for each service and model the OS handler as a case covering all the services; but this complicates the guarantee relation. Also, input arguments may be part of the state and sometimes they must not change during the execution of the event. Therefore, the relations must reflect this condition. Nevertheless, in a language not considering events,  when having sequential composition of events it is not straightforward to state in the rely and guarantee that an argument does not change during the execution of a event. Then, the specification and verification for such systems become more difficult when not having a proper framework able to deal with these features.

Actually, the concept of event has been implicitly applied in formal verification. In seL4 \cite{Klein09a}, an event is defined to wrap all system calls of the kernel, while not considering in-kernel concurrency. In the \textit{deep specification} approach \cite{Gu15}, a finite map is used to represent events of an interface or module. 
In formal verification of multicore or preemptive OS kernels \cite{Gu16,Chen16,Xu16}, rely conditions are defined as invariants to represent the environment behaviour. These studies apply the rely-guarantee technique for specific systems, rather than propose a generic framework for concurrent reactive systems. 
Event-B \cite{Abrial07} provides a refinement-based formal method for system-level modeling and analysis, in which the execution of events is not in an interleaved manner. 
In \cite{Hoang10}, Event-B is extended to mimic rely-guarantee style reasoning for concurrent programs, but not to provide a rely-guarantee framework for Event-B. 
Compositional reasoning about critical properties (e.g., safety and security) for concurrent programs has attracted many research efforts (e.g., \cite{Flan02,Mantel11,Murray16}). However, rely-guarantee-based compositional reasoning about them at system level deserves further study. For instance, noninterference of OS kernels \cite{Murray12} concerns the whole trace of events rather than the program states.

In this paper, we propose an event-based rely-guarantee reasoning approach for concurrent reactive systems. The approach supports compositional verification of functional correctness and safety as well as dealing with multicore concurrency and interruptible concurrency together. 
We introduce ``events'' \cite{Back91,Abrial07} into the rely-guarantee method to specify and verify reactive services. 
%They could be modularly composed as an event system which represents a single-processing reactive system. A set of event systems are further parallelly composed as a concurrent reactive system. 
Developers could focus on specifying and verifying reactive services. Instead, compositional specification and verification are kindly supported in our framework. 
This can offer a flexible rely-guarantee framework for modeling and verification at system level \cite{Jones15}. 
We extend the imperative language in \cite{Xu97,Nieto03} to specify imperative statements in events. %The semantics and rely-guarantee proof system of the language are thus reused. 
Other richer imperative languages, such as CSimpl \cite{Sanan17}, can also be wrapped by events for specification and verification at implementation level. This work is the first effort to study the rely-guarantee method for system-level concurrency in the literature. 

We focus on multicore concurrency and interruptible concurrency of reactive systems. First, recent multicore OS kernels such as XtratuM \cite{Carr14} and CertiKOS \cite{Gu16} are shared-variable concurrent reactive systems. Kernel instances may be executed simultaneously on different cores of a processor sharing the same memory. Interleaving may happen at arbitrary fine-grained program locations in interrupt handlers. Second, in interruptable systems, the execution of functions may be interrupted and jumps to the interrupt handler. During the execution of the handler, the function is blocked at the break point. Upon return from the handler, the system state could be substantially changed. 

In detail, the technical contributions of this work are as follows.

\begin{enumerate}
%\vspace{-2.0mm}
\item \label{contr:1} We propose an event-based language -- {\slang}, which incorporates concurrent programming and system specification languages. We define a modular composition and a parallel composition of events as well as semantics of non-deterministic occurrence and interleaving of events. ({\sectprefix} \ref{sect:pclang}) %The language could be applied to create formal specification as well as to design and implement the system ({\sectprefix} \ref{sect:pclang}). 

\item \label{contr:2} We build a rely-guarantee proof system for {\slang} and prove its soundness. We provide rules to prove properties with coarse and fine granularity. We can prove functional partial correctness of systems by rely-guarantee conditions of each event providing coarse granularity of properties, i.e. granularity at event level. We provide a compositional reasoning approach for safety properties defined as invariants providing fine granularity, i.e., granularity at internal steps of events. Invariant proof of systems can be discharged by local proof on each event. ({\sectprefix} \ref{sect:rgproof})
%We provide proof rules for both parallel composition of event systems and nondeterministic occurrence of events.  %The rely-guarantee method is widened from programming languages to system specification languages. 
%Based on the proof system, we provide compositional reasoning approach for safety in {\slang}. 
%This work is the first effort to study the rely-guarantee method for system-level concurrency in the literature. 

\item \label{contr:3} We develop two case studies: an interruptible controller for stepper motors and inter-partition communication (IPC) of ARINC 653 \cite{ARINC653p1_4} multicore kernels. We prove the functional correctness and preservation of invariants of them. ({\sectprefix} \ref{sect:study_cases})%This work is the first case of compositional reasoning for multicore kernels in the literature.

\item \label{contr:4} We develop the {\slang} language and the two case studies in Isabelle/HOL using $\approx$ 13,000 lines of new specification and proof based on those in \cite{Nieto03}. The Isabelle sources are available at \url{https://lvpgroup.github.io/picore_doc/}. 
%\item \label{contr:4} We formalize and mechanically prove the {\slang} language, rely-guarantee proof system and its soundness, compositional reasoning about safety, and the two study cases in Isabelle/HOL using $\approx$ 13,000 lines of new specification and proof based on those in \cite{Nieto03}. The Isabelle sources are available at \url{http://securify.scse.ntu.edu.sg/MicroVer/picore/}. %\footnote{The Isabelle sources include $\approx$ 2,500 lines of proof from \cite{Nieto03}.}. %All results have been mechanically proved. %We also create a concrete syntax for {\slang} which is convenient to specify and verify concurrent systems. %The sources of Isabelle/HOL are available at ``\qmark{http://securify.scse.ntu.edu.sg/...}''. 
%\vspace{-2.5mm}
\end{enumerate}

\section{The {\slang} Language}
\label{sect:pclang}
This section introduces the {\slang} language including its abstract syntax, operational semantics, and computations. We also create a concrete syntax for {\slang}, which is illustrated in our case studies. 

\zipssect
\subsection{Abstract Syntax}

%By incorporating concurrent programming and system specification languages, we create a language with four levels of elements, i.e., \emph{programs} represented by imperative languages, \emph{events} constructed using programs, \emph{event systems} comprised of events, and \emph{parallel event systems} composed by event systems. 
The abstract syntax of {\slang} is shown in {\figprefix} \ref{fig:syntax}. 
The syntax of programs representing the body of an event extends the syntax in \cite{Xu97,Nieto03} with the $\cmdnondt{r}$ command, which models nondeterminism through a state relation $r$. The $\cmdawait{\symbbexp}{\symbprog}$ command executes the body $\symbprog$ atomically whenever the boolean condition $\symbbexp$ holds. In the case study of multicore OS kernels, we use this command to model the synchronized access to shared resources from multiple partitions. As illustrated in the case study of interruptible controller, multi-level interrupts can also be modeled by this command via an interrupts stack. 
Other commands of programs are standard. 

\begin{figure}[t]
\centering
\fontsize{7pt}{0cm}
%\scriptsize
%\footnotesize
{
\footnotesize
\begin{tabular}{ll}
\textbf{Program}: & \textbf{Event}:  \vspace{2mm}\\
$
\begin{aligned}
\symbprog \ ::= & \ \cmdbasic{f} \ | \ \cmdseq{\symbprog_1}{\symbprog_2} \ | \ \cmdcond{\symbbexp}{\symbprog_1}{\symbprog_2} \\
| & \ \cmdwhile{\symbbexp}{\symbprog} \ | \ \cmdawait{\symbbexp}{\symbprog} \ | \ \cmdnondt{r} \ |  \cmdnone 
%| & \ \cmdfont{Call} \ P \ @ \ \domfont{u} 
\end{aligned}
$ \hspace{5mm}
&
$
\begin{aligned}
\symbEvt \ ::= & \ \event{\symbevtbd} & (Basic \ Event)  \\ 
 | & \ \anonevt{\symbprog} & (Inner\ of\ Event)
\end{aligned}
$ \vspace{0.3cm} \\
\textbf{Event System}: & \textbf{Parallel Event System}: \vspace{2mm} \\
$
\begin{aligned}
\symbevtsys \ ::= & \ \evtsysdef & (Event \ Set) \\
| & \ \evtseq{\symbEvt}{\symbevtsys} & (Event \ Sequence)
\end{aligned}
$\hspace{10mm}
&
$
\begin{aligned}
\symbpes \ ::= \ \parsysc  %\parsys{\mathcal{S}_1}{\mathcal{S}_2}{\mathcal{S}_n} 
\end{aligned}
$
\end{tabular}
}
\caption{Abstract Syntax of {\slang}}
\label{fig:syntax}
\zipfigaft
%\vspace{-4mm}
\end{figure}

Reaction services are modeled as events which are parameterized programs with a guard condition. The parameters indicate how a reaction service is triggered. They are the context when starting the execution of reaction services and decided by application programs, such as the parameters of an invocation to a system call. The guard condition, e.g. the interrupt flag in x86 is enabled, indicates when a reaction service can be triggered.  
The syntax for events distinguishes non-triggered events $\event{\symbevtbd}$, called basic event, from triggered events $\anonevt{\symbprog}$ that are being executed. A basic event $\event{\symbevtbd}$ is a tuple $\ell \times (g \times \symbprog)$ where $\ell$ defines the name, $g$ the guard condition, and $\symbprog$ the body of the event. $\event{(l,g,P)}$ is triggered when $g$ holds in the current state. 
Then, its body begins to be executed and the event is substituted by $\anonevt{\symbprog}$. 
We implement the parameterization of events in concrete syntax of {\slang}. %A parameterized event is defined as $\lambda (plist, \symbcore). \  \event{(l,g,P)}$, where $plist$ is a list of parameters, $plist$ and $\symbcore$ are variables in the event declaration of $\event{(l,g,P)}$. 

We define two categories of event composition in this paper: modular and parallel composition. 
An event system is a modular composition of events representing the behaviour of a single-processing system. It has two forms that we call \emph{event sequence} and \emph{event set}. The execution of an event set consists of a continuous evaluation of the guards of the events in the set. When there is an event $\event{(l,g,P)}$ in the set where $g$ holds in the current state, the event is triggered and its body $P$ executed. After $P$ finishes, the evaluation of the guards starts again looking for the next event to be executed.   
The event sequence models the sequential execution of events. 
In an event sequence  $\evtseq{(\event{\symbevtbd})}{\symbevtsys}$, when the guard condition of $\symbevtbd$ holds in the current state, $\event{\symbevtbd}$ is triggered and the event sequence transits to $\evtseq{\anonevt{\symbprog}}{\symbevtsys}$, with $\anonevt{\symbprog}$ being the body of $\alpha$. The event sequence behaves as the event system $\symbevtsys$ when the execution of $\symbprog$ finishes. This form of event sequences is able to specify the initialization of an event system. 
A concurrent reactive system is modeled by a parallel composition of event systems with shared states. It is a function from $\symbCore$ to event systems, where $\symbCore$ indicates the identifiers of event systems. This design is more general and could be applied to track executing events. For instance, we use $\symbCore$ to represent the core identifier in multicore systems. 
%Note that a model eventually terminates is not mandatory. As a matter of fact, most of the systems we study run forever.
%We introduce an auxiliary function $evts$ to query all events. 

\zipssect
\subsection{Operational Semantics}
%Since the execution of events, event systems, and parallel event systems changes the event context, a configuration $\symbconf$ of them is defined as a triple $(\symbspec, \symbstate, \symbevtctx)$, where $\symbspec$ is a specification, $\symbstate$ is a state, and $\symbevtctx : \symbCore \rightarrow \symbEvt$ is an event context. The event context indicates which event is currently executed in an event system. We use $\symbspec_\symbconf$, $\symbstate_\symbconf$, and $\symbevtctx_\symbconf$ to represent the three parts of a configuration $\symbconf$ respectively. The configuration of programs follows the traditional form of $(\symbprog, \symbstate)$ since the execution of programs does not change the event context. 

The semantics of {\slang} is defined via transition rules between configurations. We define a configuration $\symbconf$ in {\slang} as a triple  $(\symbspec, \symbstate, \symbevtctx)$ where $\symbspec$ is a specification, $\symbstate$ is a state, and $\symbevtctx : \symbCore \rightarrow \symbEvt$ is an event context. The event context indicates which event is currently executed in an event system $k$. 
%The configuration of programs follows the traditional form of $(\symbprog, \symbstate)$ since the execution of programs does not change the event context. 

A system can perform two kinds of transitions: \emph{action transitions} and \emph{environment transitions}. The former are performed by the system itself at a parallel event system or an event system; the latter by an arbitrary environment in a parallel event system, or by an event system $k_j$ when computing an event system $k_i$ with $j\neq i$. 
Transition rules for actions in events, event systems, and parallel event systems have the form $(\symbspec_1, \symbstate_1, \symbevtctx_1) \tran{\symbactk} (\symbspec_2, \symbstate_2, \symbevtctx_2)$, where $\symbactk = \actk{\symbact}{\symbcore}$ is a label indicating the kind of transition. Here $\symbact$ can be a constant to indicate a program action or an occurrence of an event $\symbEvt$. $@\symbcore$ means that the action $\symbactk$ occurs in event system $\symbcore$. %An action that occurs in event system $\symbcore$ is denoted as $\symbactk = \actk{\symbact}{\symbcore}$. 
Environment transition rules have the form $(\symbspec, \symbstate, \symbevtctx) \evtran (\symbspec, \symbstate', \symbevtctx')$. Intuitively, a transition made by the environment may change the state and the event context but not the specification.

\begin{figure}[t]
\centering
%\scriptsize
%\footnotesize
\fontsize{8pt}{0cm}

\begin{tabular}{cccc}
\myinfer{Basic}{\infer{(\cmdbasic{f}, \symbstate) \tran{c} (\cmdnone, f \ \symbstate)}{-}} \hspace{-0.3cm}
%\infer[\textsc{Basic}]{(\cmdbasic{f}, \symbstate, \symbevtctx) \tran{c} (\cmdnone, f \ \symbstate, \symbevtctx)}{-} \hspace{0.2cm}
&
\myinfer{Seq1}{\infer{(\cmdseq{\symbprog_1}{\symbprog_2}, \symbstate) \tran{\symbpcomp} (\symbprog_2, \symbstate')}{(\symbprog_1,\symbstate) \tran{\symbpcomp} (\cmdnone,\symbstate')}} \hspace{-0.3cm}
&
\myinfer{Seq2}{\infer{(\cmdseq{\symbprog_1}{\symbprog_2}, \symbstate) \tran{\symbpcomp} (\cmdseq{\symbprog_1'}{\symbprog_2}, \symbstate')}{(\symbprog_1,\symbstate) \tran{\symbpcomp} (\symbprog_1',\symbstate') \quad \symbprog_1' \neq \cmdnone}} \hspace{-0.3cm}
&
\myinfer{CondF}{\infer{(\cmdcond{\symbbexp}{\symbprog_1}{\symbprog_2}, \symbstate) \tran{\symbpcomp} (\symbprog_2, \symbstate)}{\symbstate \notin \symbbexp}}
\end{tabular}
\vspace{0.2cm}

\begin{tabular}{ccc}
\myinfer{CondT}{\infer{(\cmdcond{\symbbexp}{\symbprog_1}{\symbprog_2}, \symbstate) \tran{\symbpcomp} (\symbprog_1, \symbstate)}{\symbstate \in \symbbexp}}
&
\myinfer{WhileT}{\infer{(\cmdwhile{\symbbexp}{\symbprog}, \symbstate) \tran{\symbpcomp} (\cmdseq{\symbprog}{(\cmdwhile{\symbbexp}{\symbprog})}, \symbstate)}{\symbstate \in \symbbexp}} 
&
\myinfer{WhileF}{\infer{(\cmdwhile{\symbbexp}{\symbprog}), \symbstate) \tran{\symbpcomp} (\cmdnone, \symbstate)}{\symbstate \notin \symbbexp}}
\end{tabular}
\vspace{0.2cm}

\begin{tabular}{ccc}
\myinfer{Nondt}{\infer{(\cmdnondt{r}, \symbstate) \tran{\symbpcomp} (\cmdnone, \symbstate')}{(\symbstate,\symbstate') \in r}} \hspace{0.4cm}
&
\myinfer{Await}{\infer{(\cmdawait{\symbbexp}{\symbprog}, \symbstate) \tran{\symbpcomp} (\cmdnone, \symbstate')}{\symbstate \in \symbbexp & (\symbprog, \symbstate) \tran{\symbpcomp^{*}} (\cmdnone, \symbstate')}} \hspace{0.4cm}
&
\myinfer{InnerEvt}{\infer{(\anonevt{\symbprog},\symbstate,\symbevtctx) \trank{\symbpcomp}{k} (\anonevt{\symbprog'},\symbstate',\symbevtctx)}{(\symbprog,\symbstate) \tran{\symbpcomp} (\symbprog',\symbstate')}}
\end{tabular}
\vspace{0.1cm}

\begin{tabular}{cc}
\myinfer{BasicEvt}{\infer{(\event{\symbevtbd}, \symbstate, \symbevtctx) \trank{\event{\symbevtbd}}{\symbcore} (\anonevt{\symbprog}, \symbstate, \symbevtctx')}{
%\begin{tabular}{l}
%$\symbprog = body(\symbevtbd) \quad \symbstate \in guard(\symbevtbd)$ \\
%$\symbevtctx' = \symbevtctx(k \mapsto \event{\symbevtbd})$
%\end{tabular}
\symbprog = body(\symbevtbd) \quad \symbstate \in guard(\symbevtbd) \quad \symbevtctx' = \symbevtctx(k \mapsto \event{\symbevtbd})
}} \hspace{0.5cm}
& 
\myinfer{EvtSet}{\infer{(\evtsysdef, \symbstate, \symbevtctx) \trank{\symbEvt_i}{\symbcore} (\evtseq{\symbEvt_i'}{\evtsysdef},\symbstate, \symbevtctx')}{i \leq n & (\symbEvt_i, \symbstate, \symbevtctx) \trank{\symbEvt_i}{\symbcore} (\symbEvt_i',\symbstate, \symbevtctx')}}
\end{tabular}
\vspace{0.1cm}

\begin{tabular}{ccc}
\myinfer{EvtSeq1}{\infer{(\evtseq{\symbEvt}{\symbevtsys}, \symbstate, \symbevtctx) \trank{\symbact}{\symbcore} (\evtseq{\symbEvt'}
{\symbevtsys}, \symbstate', \symbevtctx')}{(\symbEvt, \symbstate, \symbevtctx) \trank{\symbact}{\symbcore} (\symbEvt', \symbstate', \symbevtctx') \quad \symbEvt' \neq \anonevt{\cmdnone}}} \hspace{-5mm}
&
\myinfer{EvtSeq2}{\infer{(\evtseq{\symbEvt}{\symbevtsys}, \symbstate, \symbevtctx) \trank{\symbact}{\symbcore} (\symbevtsys, \symbstate', \symbevtctx')}
{(\symbEvt, \symbstate, \symbevtctx) \trank{\symbact}{\symbcore} (\anonevt{\cmdnone}, \symbstate', \symbevtctx')}} \hspace{-5mm}
&
\myinfer{Par}{\infer{(\symbpes, \symbstate, \symbevtctx) \trank{\symbact}{\symbcore} (\symbpes', \symbstate', \symbevtctx')}{
%\begin{tabular}{l}
%$(\symbpes(\symbcore), \symbstate, \symbevtctx) \trank{\symbact}{\symbcore} (\symbevtsys', \symbstate', \symbevtctx')$ \\
%$\symbpes' = \symbpes(\symbcore \mapsto \symbevtsys')$
%\end{tabular}
(\symbpes(\symbcore), \symbstate, \symbevtctx) \trank{\symbact}{\symbcore} (\symbevtsys', \symbstate', \symbevtctx') \quad \symbpes' = \symbpes(\symbcore \mapsto \symbevtsys')
}}
\end{tabular}
\caption{Operational Semantics of {\slang}}
\label{fig:semantics}
\zipfigaft
\end{figure}

Transition rules of {\slang} are shown in {\figprefix} \ref{fig:semantics}. Transition rules of programs follow the traditional form of $(\symbprog_1, \symbstate_1) \tran{c} (\symbprog_2, \symbstate_2)$, since the execution of programs does not change the event context. 
 $\tran{\symbpcomp^{*}}$ in the $\textsc{Await}$ rule is the reflexive transitive closure of $\tran{\symbpcomp}$. $\textsc{Nondt}$ transits from a state $s$ to state $s'$ if $(s,s')\in r$ and it blocks otherwise. Other transition rules of programs are standard. 
 
The execution of $\anonevt{\symbprog}$ mimics program $\symbprog$. The $\textsc{BasicEvt}$ rule shows the occurrence of an event when its guard is true in the current state. It updates the specification and context of the current state to the program bounded to the triggered event and the event itself, respectively. 
The $\textsc{EvtSet}$, $\textsc{EvtSeq1}$, and $\textsc{EvtSeq2}$ rules specify the occurrence and execution of events in an event set. After the execution of the event, the event system behaves as the original event set. 
%Note that a model eventually terminates is not mandatory. As a matter of fact, most of the systems we study run forever.

The $\textsc{Par}$ rule shows that the execution of a parallel event system is modeled by a non-deterministic interleaving of event systems. $\symbpes(\symbcore \mapsto \symbevtsys')$ updates the function $\symbpes$ using $\symbevtsys'$ to replace the mapping of $\symbcore$.
The parallel composition of event systems is fine-grained since small steps in events are interleaved in the semantics of {\slang}. This design relaxes the atomicity of events in other approaches (e.g., Event-B \cite{Abrial07}).

Our framework in this paper tackles with partial correctness and therefore we are assuming program termination. In the semantics, state transformation of $\cmdbasic{f}$ is atomic as well as guard evaluation of $\cmdfont{Cond}$, $\cmdfont{While}$, and $\cmdfont{Await}$ statements. It is reasonable for specification for two reasons. First, complicated guard conditions in programming languages may be decomposed and specified by introducing local variables in {\slang}. Second, shared variables in concurrent programs, such as multicore OS kernels, are usually controlled by mutex.

\zipssect
\subsection{Computation}

A \emph{computation} of {\slang} is a sequence of transitions. 
We define the set of computations of parallel event systems as $\compfun(\symbpes)$, which is a set of lists of configurations inductively defined as follows. The singleton list is always a computation. Two consecutive configurations are part of a computation if they are the initial and final configurations of an environment or action transition. The operator $\#$ in $e\# l$ represents the insertion of element $e$ in list $l$.

\vspace{-4mm}
\[\fontsize{8pt}{0cm}
\begin{aligned}
& \cmdfont{One}: [(\symbpes, \symbstate, \symbevtctx)] \in  \compfun(\symbpes) \\
& \cmdfont{Env}:(\symbpes, \symbstate_1, \symbevtctx_1)\#cs \in  \compfun(\symbpes)  \Longrightarrow (\symbpes, \symbstate_2, \symbevtctx_2)\#(\symbpes, \symbstate_1, \symbevtctx_1)\#cs \in  \compfun(\symbpes) \\
& \cmdfont{Act}: (\symbpes_2, \symbstate_2, \symbevtctx_2) \tran{\symbactk} (\symbpes_1, \symbstate_1, \symbevtctx_1) \wedge (\symbpes_1, \symbstate_1, \symbevtctx_1)\#cs \in  \compfun(\symbpes) \Longrightarrow (\symbpes_2, \symbstate_2, \symbevtctx_2)\#(\symbpes_1, \symbstate_1, \symbevtctx_1)\#cs \in  \compfun(\symbpes)
\end{aligned}
\]
%\vspace{-4mm}
%\[%\footnotesize
%\left\{
%\begin{aligned}
%& [(\symbpes, \symbstate, \symbevtctx)] \in  \compfun(\symbpes) \\
%& (\symbpes, \symbstate_1, \symbevtctx_1)\#cs \in  \compfun(\symbpes)  \Longrightarrow (\symbpes, \symbstate_2, \symbevtctx_2)\#(\symbpes, \symbstate_1, \symbevtctx_1)\#cs \in  \compfun(\symbpes) \\
%& (\symbpes_2, \symbstate_2, \symbevtctx_2) \tran{\symbactk} (\symbpes_1, \symbstate_1, \symbevtctx_1) \wedge (\symbpes_1, \symbstate_1, \symbevtctx_1)\#cs \in  \compfun(\symbpes) \\
%& \quad \quad \quad \Longrightarrow (\symbpes_2, \symbstate_2, \symbevtctx_2)\#(\symbpes_1, \symbstate_1, \symbevtctx_1)\#cs \in  \compfun(\symbpes)
%\end{aligned}
%\right.
%\]
%\vspace{-4mm}

The computations of programs, events, and event systems are defined in a similar way. We use $\compfun(\symbpes)$ to denote the set of computations of a parallel event system $\symbpes$. The function $\compfun(\symbpes, \symbstate, \symbevtctx)$ denotes the computations of $\symbpes$ executing from an initial state $\symbstate$ and event context $\symbevtctx$. The computations of programs, events, and event systems are also denoted as the $\compfun$ function. 
We say that a parallel event system $\symbpes$ is a \emph{closed system} when there is no environment transition in computations of $\symbpes$. 
%or the environment transitions do not change the state and event context. 

\section{Compositional Verification of Functional Correctness and Safety}
\label{sect:rgproof}
For the purpose of compositional reasoning, we propose a rely-guarantee proof system for {\slang} in this section. 
We first introduce the rely-guarantee specification and its validity. Then, we present a set of proof rules and their soundness for compositionality, and compositional reasoning about safety properties. 

Formal specifications and proofs in existing rely-guarantee methods only consider the state of programs and they focus on traditional imperative sequential languages. At event-system level, events are basic components for the specification and compositional reasoning is conducted using the rely-guarantee conditions of events. This decomposition allows us to ease the specification in the rely and the guarantee of local variables to events, as well as the reasoning on sequences of events and properties involving local variables. %An event-based specification provides better granularity than traditional rely-guarantee methods for imperative programs. 
We consider the verification of two different kinds of properties in the rely-guarantee proof system for reactive systems: pre- and post-conditions of events and invariants. We use the former for the verification of functional correctness, whilst we use the latter on the verification of safety properties concerning the internal steps of events. For instance, in the case of the interruptible controller, a safety property is that collisions must not happen even during internal steps of the \textit{forward} and \textit{backward} system calls.
Other critical properties can also be defined considering the execution trace of events, e.g. information-flow security \cite{Mantel11,Murray16}. 

\zipssect
\subsection{Rely-Guarantee Specification}

A rely-guarantee specification for a system is a quadruple $RGCond = \rgconddefault$, where $pre$ is the pre-condition, $R$ is the rely condition, $G$ is the guarantee condition, and $pst$ is the post-condition. The assumption and commitment functions are denoted by $A$ and $C$ respectively. For each computation $\symbcomp \in \compfun(\symbpes)$, we use $\symbcomp_i$ to denote the configuration at index $i$. For convenience, we use $\symbcomp$ to denote computations of programs, events, and event systems.  $\symbspec_{\symbcomp_i}$, $\symbstate_{\symbcomp_i}$, and $\symbevtctx_{\symbcomp_i}$ represent the elements of $\symbcomp_i =(\symbspec, \symbstate, \symbevtctx)$.

\vspace{-3mm}
\begin{equation*}\fontsize{8pt}{0cm}
\begin{aligned}
& \assumefun(pre, R) \equiv \{\symbcomp \mid \symbstate_{\symbcomp_0} \in pre \wedge (\forall i < len(\symbcomp) - 1. \ (\symbcomp_i \evtran \symbcomp_{i+1}) \longrightarrow (\symbstate_{\symbcomp_i},\symbstate_{\symbcomp_{i+1}}) \in R)\} \\
& \commitfun(G, pst) \equiv \{\symbcomp \mid \ (\forall i < len(\symbcomp) - 1. \ (\symbcomp_i \tran{\symbactk} \symbcomp_{i+1}) \longrightarrow (\symbstate_{\symbcomp_i},\symbstate_{\symbcomp_{i+1}}) \in G) \wedge (\symbspec_{last(\symbcomp)} = \cmdnone \longrightarrow \symbstate_{\symbcomp_n} \in pst)\} 
\end{aligned}
\end{equation*}

%For an event, the commitment function is similar, but the condition $\symbspec_{last(\symbcomp)} = \anonevt{\cmdnone}$. Since event systems and parallel event systems execute forever, we release the post-condition in the commitment function of them as follows. . 
%
%\vspace{-0.5cm}
%\begin{equation*}\footnotesize
%\begin{aligned}
%\commitfun(G, pst) \equiv \{\symbcomp \mid (\forall i < len(\symbcomp) - 1. \ (\symbcomp_i \tran{\symbactk} \symbcomp_{i+1}) \longrightarrow (\symbstate_{\symbcomp_i},\symbstate_{\symbcomp_{i+1}}) \in G)\}
%\end{aligned}
%\end{equation*}

We define the validity of a rely-guarantee specification  $\rgconddefault$ in a parallel event system as follows.

\vspace{-3mm}
%\vspace{-0.5cm}
\[
\RGSAT{\symbpes}{\rgcond{pre}{R}{G}{pst}} \equiv \forall \symbstate, \symbevtctx. \ \compfun(\symbpes, \symbstate, \symbevtctx) \cap \assumefun(pre, R) \subseteq \commitfun(G, pst)
\]
%\vspace{-0.5cm}

Intuitively, validity represents that the set of computations starting at configuration $(\symbpes, \symbstate, \symbevtctx)$, with $\symbstate \in pre$ and any environment transition belonging to the rely relation $R$, is a subset of the set of computations where action transitions belongs to the guarantee relation $G$ and where if a system terminates, then the final states belongs to $pst$. 
Validity for programs, events, and event systems are defined in a similar way.

\zipssect
\subsection{Compositional Proof Rules}
We present the proof rules in {\figprefix} \ref{fig:proofrule}, which gives us a relational proof method for concurrent systems. $UNIV$ is the universal set. We first define $stable(f,g) \equiv \forall x,y. \ x \in f \wedge (x, y) \in g \longrightarrow y \in f$. Thus, $stable(pre, rely)$ means that the pre-condition is stable when the rely condition holds.

\begin{figure}
\centering
%\scriptsize
%\footnotesize
\fontsize{8pt}{0cm}

\begin{tabular}{ccc}
\myinfer{Basic}{\infer{\rgsat{(\cmdbasic{f})}{\rgconddefault}}
{
\begin{tabular}{l}
$pre \subseteq \{\symbstate \mid f(\symbstate) \in pst\}$ \\ 
$\{(\symbstate, \symbstate') \mid \symbstate \in pre \wedge \symbstate' = f(\symbstate)\} \in G $ \\
$stable(pre, R) \quad stable(pst, R)$
\end{tabular}
}} \hspace{-0.5cm}
&
\myinfer{Cond}{\infer{\rgsat{(\cmdcond{\symbbexp}{P_1}{P_2})}{\rgconddefault}}
{
\begin{tabular}{l}
$\rgsat{P_1}{\rgcond{pre \cap \symbbexp}{R}{G}{pst}}$ \\ 
$\rgsat{P_2}{\rgcond{pre \cap - \symbbexp}{R}{G}{pst}}$ \\
$stable(pre, R) \quad \forall \symbstate. \ (\symbstate, \symbstate) \in G$
\end{tabular}
}} \hspace{-0.5cm}
&\myinfer{Seq}{\infer{\rgsat{(\cmdseq{P}{Q})}{\rgconddefault}}
{
\begin{tabular}{l}
$\rgsat{P}{\rgcond{pre}{R}{G}{m}}$ \\
$\rgsat{Q}{\rgcond{m}{R}{G}{pst}}$
\end{tabular}
%\rgsat{P}{\rgcond{pre}{R}{G}{m}} & \rgsat{Q}{\rgcond{m}{R}{G}{pst}} 
}}
\end{tabular}
%\vspace{0.3cm}

\begin{tabular}{cc}
\myinfer{While}{\infer{\rgsat{(\cmdwhile{\symbbexp}{P})}{\rgconddefault}}
{
\begin{tabular}{l}
$\rgsat{P}{\rgcond{pre \cap \symbbexp}{R}{G}{pre}} \quad pre \cap - \symbbexp \subseteq pst$ \\
$stable(pre, R) \quad stable(pst, R) \quad \forall \symbstate. \ (\symbstate, \symbstate) \in G $
\end{tabular}
}}
&
\myinfer{Await}{\infer{\rgsat{(\cmdawait{\symbbexp}{\symbprog})}{\rgconddefault}}
{
\begin{tabular}{l}
$\forall V.\ \rgsat{P}{\rgcond{pre \cap \symbbexp \cap \{V\}}{Id}{UNIV}{\{\symbstate \mid (V, \symbstate) \in G\} \cap pst}}$ \\
$stable(pre, R) \quad stable(pst, R)$
\end{tabular}
}}
\end{tabular}
%\vspace{0.3cm}

\begin{tabular}{cc}
\myinfer{Nondt}{\infer{\rgsat{(\cmdnondt{r})}{\rgconddefault}}
{
\begin{tabular}{l}
$pre \subseteq \{s \mid \ (\forall \symbstate'. \ (\symbstate, \symbstate') \in r \longrightarrow \symbstate' \in pst) \wedge (\exists \symbstate'. \ (\symbstate, \symbstate') \in r)\}$ 
\\
$\{(\symbstate, \symbstate') \mid \symbstate \in pre \wedge (\symbstate, \symbstate') \in r\} \subseteq G \quad stable(pre, R) \quad stable(pst, R)$
\end{tabular}
}}
&
\myinfer{Conseq}{\infer{\rgsat{\symbspec}{\rgconddefault}}
{
\begin{tabular}{l}
$pre \subseteq pre' \; R \subseteq R' \;\; G' \subseteq G \;\; pst' \subseteq pst$ \\
$\rgsat{\symbspec}{\rgcond{pre'}{R'}{G'}{pst'}}$ 
\end{tabular}
}} 
\end{tabular}
%\vspace{0.1cm}

\begin{tabular}{ccc}
\myinfer{BasicEvt}{\infer{\rgsat{\event{\symbevtbd}}{\rgconddefault}}
{
\begin{tabular}{l}
$\rgsat{body(\symbevtbd)}{\rgcond{pre \cap guard(\symbevtbd)}{R}{G}{pst}}$\\
$stable(pre, R) \quad \forall \symbstate. \ (\symbstate, \symbstate) \in G$
\end{tabular}
}} \hspace{-0.6cm}
&
\myinfer{Inner}{\infer{\rgsat{(\anonevt{P})}{\rgconddefault}}
{\rgsat{P}{\rgconddefault}}} \hspace{-0.6cm}
&
\myinfer{EvtSeq}{\infer{\rgsat{(\evtseq{\symbEvt}{\symbevtsys})}{\rgconddefault}}
{
\begin{tabular}{l}
$\rgsat{\symbEvt}{\rgcond{pre}{R}{G}{m}}$ \\
$\rgsat{\symbevtsys}{\rgcond{m}{R}{G}{pst}}$
\end{tabular}
%\rgsat{\symbEvt}{\rgcond{pre}{R}{G}{m}} \quad \rgsat{\symbevtsys}{\rgcond{m}{R}{G}{pst}}
}} 
\end{tabular}
%\vspace{0.1cm}

\begin{tabular}{cc}
\myinfer{EvtSet}{\infer{\rgsat{(\evtsysdef)}{\rgconddefault}}
{
\begin{tabular}{l}
$\forall i \leq n. \ \rgsat{\symbEvt_i}{\rgcond{pres_i}{Rs_i}{Gs_i}{psts_i}}$ \\
$ stable(pre, R)\quad \forall i, j \leq n. \ psts_i \subseteq pres_j $\\
$\forall i \leq n. \ pre \subseteq pres_i \quad \forall i \leq n. \ psts_i \subseteq pst$\\
$\forall i \leq n. \ R \subseteq Rs_i \quad \forall i \leq n. \ Gs_i \subseteq G  \quad \forall \symbstate. \ (\symbstate, \symbstate) \in G$
\end{tabular}
}}
&
\myinfer{Par}{\infer{\rgsat{\symbpes}{\rgconddefault}}
{
\begin{tabular}{l}
$\forall \symbcore. \ \rgsat{\symbpes(\symbcore)}{\rgcond{pres_\symbcore}{Rs_\symbcore}{Gs_\symbcore}{psts_\symbcore}}$ \\
$\forall \symbcore. \ pre \subseteq pres_\symbcore \quad \forall \symbcore. \ psts_\symbcore \subseteq pst$ \\
$ \forall \symbcore. \ Gs_\symbcore \subseteq G \quad \forall \symbcore. \ R \subseteq Rs_\symbcore $ \\
$\forall \symbcore, \symbcore'. \ \symbcore \neq \symbcore' \longrightarrow Gs_\symbcore \subseteq Rs_{\symbcore'}$
\end{tabular}
}}
\end{tabular}

\caption{Rely-guarantee Proof Rules for {\slang}}
\label{fig:proofrule}
\zipfigaft
\end{figure}

For $\cmdawait{\symbbexp}{\symbprog}$, by the semantics of the command, the evaluation of the condition $\symbbexp$ and the execution of the body $\symbprog$ are done atomically. Thus, the state transition of the command must satisfy the guarantee condition. This is presented in the pre- and post-conditions of $\symbprog$ in the assumptions of the $\textsc{Await}$ rule. We use an universally quantified variable $V$ to relate the state before and after the transformation. The intermediate state changes during the execution of $\symbprog$ must not guarantee anything, thus the guarantee condition is $UNIV$. Since $\symbprog$ is executed atomically, the environment cannot change the state, i.e. the rely condition is the identity relation $Id$. To ensure that both the pre- and post-conditions are preserved after environment transitions we request stability of $pre$ and $pst$. 

For $\cmdnondt{r}$, any state change in $r$ requires that $pst$ holds immediately after the action transition and the transition should be in $G$ relation. Before and after the action transition there may be a number of environment transitions that can modify the state. To ensure that both the pre- and post-conditions are preserved after environment transitions we request stability of $pre$ and $pst$. 
Other proof rules for programs are standard, and we reuse them from \cite{Nieto03}. 
%Stability is defined as follows.

%\vspace{-0.2cm}
%\[
%stable(f,g) \equiv \forall x,y. \ x \in f \wedge (x, y) \in g \longrightarrow y \in f
%\]
%\vspace{-0.5cm}

For an inner of events, it is just a wrapper of a program, and they have the same state and event context in their computations according to the $\textsc{InnerEvt}$ transition rule in {\figprefix} \ref{fig:semantics}. Therefore, $\anonevt{P}$ satisfies the rely-guarantee specification iff the program $P$ satisfies the specification. For a basic event, it satisfies the rely-guarantee specification, if its body satisfies the rely-guarantee condition with an augmented pre-condition with the guard condition of the event. Since the occurrence of an event does not change the state ($\textsc{BasicEvt}$ rule in {\figprefix} \ref{fig:semantics}), we require that $\forall \symbstate. \ (\symbstate, \symbstate) \in G$, i.e. $Id \subseteq G$. As illustrated in our case studies, it is reasonable since the $Id$ set is usually a subset of the rely condition of other event systems. Moreover, there may be a number of environment transitions before the event occurs. $stable(pre, R)$ ensures that $pre$ holds during the environment transitions. 

Regarding the proof rules for event systems, sequential composition of events is modeled by rule $\textsc{EvtSeq}$, which is similar to the rule for sequential statement. 
%Recall that when an event occurs in an event set, the event executes until it finishes in the event system. Then, the event system behaves as the event set. Thus, events in an event system do not execute in interleaving manner. 
In order to prove that an event set satisfies its rely-guarantee specification, we have to prove eight premises ($\textsc{EvtSet}$ rule in {\figprefix} \ref{fig:proofrule}). It is necessary that each event together with its specification is derivable in the system (Premise 1). Since the event set behaves as itself after an event finishes, then the post-condition of each event should imply the pre-condition of each event (Premise 2), and the pre-condition for the event set has to imply the pre-conditions of all events (Premise 3). An environment transition for event $i$ corresponds to a transition from the environment of the event set (Premise 4). The guarantee condition $Gs_i$ of each event must be in the guarantee condition of the event set, since an action transition of the event set is performed by one of its events (Premise 5). The post-condition of each event must be in the overall post-condition (Premise 6). The last two refer to stability of the pre-condition and identity of the guarantee relation.

The $\textsc{Conseq}$ rule can be applied to programs, events, and event systems, where the specification is denoted as $\symbspec$, allowing us to strengthen the assumptions and weaken the commitments. 

We now introduce the proof rule $\textsc{Par}$ for parallel composition of event systems. In order to prove that a concurrent reactive system satisfies its rely-guarantee specification, we have to prove six premises ($\textsc{Par}$ rule in {\figprefix} \ref{fig:proofrule}). A concurrent system in {\slang} is modeled as a function from $\symbCore$ to event systems. It is necessary that each event system $\symbpes(\symbcore)$ satisfies its specification $\rgcond{pres_\symbcore}{Rs_\symbcore}{Gs_\symbcore}{psts_\symbcore}$ (Premise 1). The pre-condition for the parallel composition imply all the event system's pre-conditions (Premise 2). The overall post-condition must
be a logical consequence of all post-conditions of event systems (Premise 3). 
Since an action transition of the concurrent system is performed by one of its event system, the guarantee condition $Gs_\symbcore$ of each event system must be a subset of the overall guarantee condition $G$ (Premise 4). 
An environment transition $Rs_\symbcore$ for the event system $\symbcore$ corresponds to a transition from the overall environment $R$ (Premise 5).
An action transition of an event system $\symbcore$ should be defined in the rely condition of another event system $\symbcore'$, where $\symbcore \neq \symbcore'$ (Premise 6).

Besides the proof rule for each language constructor of {\slang} ({\figprefix} \ref{fig:proofrule}), we also define a set of auxiliary proof rules for programs to ease complicated proof of large programs, as shown in {\figprefix} \ref{fig:auxproofrule}. The first two rules shows the union of pre-conditions and intersection of post-conditions. The $\textsc{UnivPre}$ rule is usefull for $\cmdawait{\symbbexp}{\symbprog}$ command, due to the first premise of the $\textsc{Await}$ rule in {\figprefix} \ref{fig:proofrule}. The last rule $\textsc{EmptyPre}$ means that any program $P$ satisfies a rely-guarantee specification with an empty pre-condition. It is usually applied for conditional statements when the condition is $false$. 

\begin{figure}
\centering
%\scriptsize
%\footnotesize
\fontsize{8pt}{0cm}

\begin{tabular}{cccc}
\myinfer{UnPre}{\infer{\rgsat{P}{\rgcond{pre \cup pre'}{R}{G}{pst}}}
{
\begin{tabular}{l}
$\rgsat{P}{\rgcond{pre}{R}{G}{pst}}$ \\ 
$\rgsat{P}{\rgcond{pre'}{R}{G}{pst}}$ 
\end{tabular}
}} \hspace{-0.5cm}
&
\myinfer{IntPost}{\infer{\rgsat{P}{\rgcond{pre}{R}{G}{pst \cap pst'}}}
{
\begin{tabular}{l}
$\rgsat{P}{\rgcond{pre}{R}{G}{pst}}$ \\ 
$\rgsat{P}{\rgcond{pre}{R}{G}{pst'}}$ 
\end{tabular}
}} \hspace{-0.5cm}
&
\myinfer{UnivPre}{\infer{\rgsat{P}{\rgcond{pre}{R}{G}{pst}}}
{
\begin{tabular}{l}
$\forall v \in pre.$ \\
$\quad \rgsat{P}{\rgcond{\{v\}}{R}{G}{pst}}$
\end{tabular}
}} \hspace{-0.5cm}
&
\myinfer{EmptyPre}{\infer{\rgsat{P}{\rgcond{\{\}}{R}{G}{pst}}}
{ - }} \hspace{-0.5cm}
\end{tabular}

\caption{Auxiliary Rely-guarantee Proof Rules for Programs}
\label{fig:auxproofrule}
\zipfigaft
\end{figure}

\subsection{Soundness of Proof System}
\label{subsect:sound_proof}
Finally, the soundness theorem for $\symbspec$ being a specification of programs, events, event systems, or parallel event systems, relates specifications proven on the proof system with its validity. 

\begin{theorem}[\rm Soundness] If \ $\rgsat{\symbspec}{\rgconddefault}$, then $\RGSAT{\symbspec}{\rgconddefault}$.
\end{theorem}

The soundness of rules for programs is discussed in detail in \cite{Xu97} and we reuse the Isabelle/HOL sources of \cite{Nieto03}. The soundness of auxiliary proof rules for programs are straightforward and proved by induction on rules of programs constructors. 
The soundness of rules for events is obvious and proved by the rules for programs. 

To prove soundness of rules for event systems, we first show how to decompose a computation of event systems into computations of its events. 
We define an equivalent relation on computations as follows. Here, we concern the state, event context, and transitions, but not the specification of a configuration. 

\begin{definition}[Simulation of Computations]
A computation $\symbcomp_1$ is a simulation of $\symbcomp_2$, denoted as $\compsim{\symbcomp_1}{\symbcomp_2}$, if $len(\symbcomp_1) = len(\symbcomp_2)$ and $\forall i < len(\symbcomp_1) - 1. \ \symbstate_{\symbcomp_{1_i}} = \symbstate_{\symbcomp_{2_i}} \wedge \symbevtctx_{\symbcomp_{1_i}} = \symbevtctx_{\symbcomp_{2_i}} \wedge (\symbcomp_{1_i} \tran{\symbactk} \symbcomp_{1_{i+1}}) = (\symbcomp_{2_i} \tran{\symbactk} \symbcomp_{2_{i+1}})$.
%\begin{itemize}
%\item $len(\symbcomp_1) = len(\symbcomp_2)$
%\item $\forall i < len(\symbcomp_1) - 1. \ \symbstate_{\symbcomp_{1_i}} = \symbstate_{\symbcomp_{2_i}} \wedge \symbevtctx_{\symbcomp_{1_i}} = \symbevtctx_{\symbcomp_{2_i}} \wedge (\symbcomp_{1_i} \tran{\symbactk} \symbcomp_{1_{i+1}}) = (\symbcomp_{2_i} \tran{\symbactk} \symbcomp_{2_{i+1}})$
%\end{itemize}
\end{definition}

In order to decompose computations of event systems to those of events, we define serialization of events based on the simulation of computations.  

\begin{definition}[Serialization of Events]
A computation $\symbcomp$ of event systems is a serialization of a set of events $\{\symbEvt_1, \symbEvt_2, ..., \symbEvt_n\}$, denoted by $\Serialize{\symbcomp}{\{\symbEvt_1, \symbEvt_2, ..., \symbEvt_n\}}$, iff there exist a set of computations $\symbcomp_1, ..., \symbcomp_m$, where for $1 \leq i \leq m$ there exists $1 \leq k \leq n$ that $\symbcomp_i \in \compste(\symbEvt_k)$, such that $\compsim{\symbcomp}{\symbcomp_1 \# \symbcomp_2 \# ... \# \symbcomp_m}$.
\end{definition}

Then, we can decompose a computation of an event system into a set of computation of its events as follows. 

\begin{lemma}
\label{lm:seri}
For any computation $\symbcomp$ of an event system $\symbevtsys$, $\Serialize{\symbcomp}{evts(\symbevtsys)}$. 
\end{lemma}

The soundness of the $\textsc{EvtSeq}$ rule is proved by two cases. For any computation $\symbcomp$ of ``$\evtseq{\symbEvt}{\symbevtsys}$'', the first case is that the execution of event $\symbEvt$ does not finish in $\symbcomp$. In such a case, $\Serialize{\symbcomp}{\{\symbEvt\}}$. By the first premise of this rule, we can prove the soundness. In the second case, the execution of event $\symbEvt$ finishes in $\symbcomp$. In such a case, we have $\symbcomp = \symbcomp_1 \# \symbcomp_2$, where $\Serialize{\symbcomp_1}{\{\symbEvt\}}$ and $\Serialize{\symbcomp_2}{evts(\symbevtsys)}$. By the two premises of this rule, we can prove the soundness. 
The soundness of the $\textsc{EvtSet}$ rule is complicated. 
From {\lemmaprefix} \ref{lm:seri}, we have that for any computation $\symbcomp$ of the event set, $\compsim{\symbcomp}{\symbcomp_1 \# \symbcomp_2 \# ... \# \symbcomp_m}$, for $1 \leq i \leq m$ there exists $1 \leq k \leq n$ that $\symbcomp_i \in \compste(\symbEvt_k)$. 
When $\symbcomp$ is in $\assumefun(pre, R)$, from $\forall i \leq n, j \leq n. \ psts_i \subseteq pres_j$, $\forall i \leq n. \ pre \subseteq pres_i$, and $\forall i \leq n. \ R \subseteq Rs_i$, we have that there is one $k$ for each $\symbcomp_i$ that $\symbcomp_i$ is in $\assumefun(pres_k, Rs_k)$. By the first premise in the $\textsc{EvtSet}$ rule, we have $\symbcomp_i$ is in $\commitfun(Gs_k, psts_k)$. Finally, with $\forall i \leq n. \ Gs_i \subseteq G$ and $\forall i \leq n. \ psts_i \subseteq pst$, we have that $\symbcomp$ is in $\commitfun(G, pst)$.

To prove the soundness of the $\textsc{PAR}$ rule, we first use \emph{conjoin} in {\defprefix} \ref{def:conjoin} to decompose a computation of parallel event systems into computations of its event systems. Computations of a set of event systems can be combined into a computation of the parallel composition of them, iff they have the same state and event context sequences, as well as not having the action transition at the same time. The resulting computation of $\symbpes$ also has the same state and event context sequences. Furthermore, in this computation a transition is an action transition on core $\symbcore$ if this is the action in the computation of event system $\symbcore$ at the corresponding position; a transition is an environment transition if this is the case in all computations of event systems at the corresponding position. 
By the definition, we have that the semantics is compositional as shown in {\lemmaprefix} \ref{lm:semaitc_compositional}. Then, the soundness of the $\textsc{Par}$ rule is proved by a similar way in \cite{Xu97,Nieto03}.

\begin{definition}
\label{def:conjoin}
A computation $\symbcomp$ of a parallel event system $\symbpes$ and a set of computations $\symbcompk: \symbCore \rightarrow \compstes$ conjoin, denoted by $\compconjoin{\symbcomp}{\symbcompk}$, iff
\begin{itemize}
\item $\forall \symbcore. \ len(\symbcomp) = len(\symbcompk(\symbcore))$. 
\item $\forall \symbcore, j < len(\symbcomp). \ \symbstate_{\symbcomp_j} = \symbstate_{\symbcompk(\symbcore)_j} \wedge \symbevtctx_{\symbcomp_j} = \symbevtctx_{\symbcompk(\symbcore)_j}$.
\item $\forall \symbcore, j < len(\symbcomp). \ \symbspec_{\symbcomp_j}(\symbcore) = \symbspec_{\symbcompk(\symbcore)_j}$. 
\item for any $j < len(\symbcomp) - 1$, one of the following two cases holds:
	\begin{itemize}
	\item $\symbcomp_j \evtran \symbcomp_{j+1}$, and $\forall \symbcore. \ \symbcompk(\symbcore)_j \evtran \symbcompk(\symbcore)_{j+1}$. 
	\item $\symbcomp_j \trank{\symbact}{\symbcore_1} \symbcomp_{j+1}$, $\symbcompk(\symbcore_1)_j \trank{\symbact}{\symbcore_1} \symbcompk(\symbcore_1)_{j+1}$, and $\forall \symbcore \neq \symbcore_1. \ \symbcompk(\symbcore)_j \evtran \symbcompk(\symbcore)_{j+1}$. 
	\end{itemize}
\end{itemize}
\end{definition}

\begin{lemma}
\label{lm:semaitc_compositional}
The semantics of {\slang} is compositional, i.e., $\compfun(\symbpes, \symbstate, \symbevtctx) = \{\symbcomp \mid (\exists \symbcompk \mid (\forall \symbcore. \ \symbcompk(\symbcore) \in \compfun(\symbpes(\symbcore), \symbstate, \symbevtctx)) \wedge \compconjoin{\symbcomp}{\symbcompk})\}$. 
\end{lemma}

\zipssect
\subsection{Safety Verification}
%\label{sect:invar}

This subsection discusses formal verification of safety properties defined as invariants preserved in each internal transition in an event. 
%\qmark{Functional correctness of systems is proved by considering pre- and post-conditions of events. However, it is necessary to reason safety by considering the internal steps of events. For instance, in the case of the interruptible controller, collision must not happen even during internal steps of the \textit{forward} and \textit{backward} system calls. }
We use a set of states $Init$ to describe the possible initial states of a parallel event system $\symbpes$, and we say that a set of states $I$ is an invariant of $\symbpes$ with respect to $Init$ if for each reachable state $\symbstate$ from an initial state in $Init$, $\symbstate \in I$. We regard a parallel event system as a closed system for safety, i.e., it has no environment transition ($no\_envt(\symbcomp)$). 

A set of states $I$ is an invariant of $\symbpes$ w.r.t. $Init$, iff 

\vspace{-3mm}
$$%\fontsize{8pt}{0cm}
\forall \symbstate, \symbevtctx, \symbcomp. \ \symbstate \in Init \wedge \symbcomp \in \compfun(\symbpes, \symbstate, \symbevtctx) \wedge no\_envt(\symbcomp) \longrightarrow (\forall i < len(\symbcomp). \ \symbstate_{\symbcomp_i} \in I)
$$

To show that $I$ is an invariant of $\symbpes$, it suffices to show that (1) $I$ initially holds in $Init$, and (2) $I$ is preserved by each transition of $\symbpes$. First, we prove that each event satisfies its rely-guarantee specification. Then, by the $\textsc{EvtSet}$ and $\textsc{EvtSeq}$ proof rules in {\figprefix} \ref{fig:proofrule}, we can get the rely-guarantee specification for each event system of $\symbpes$, i.e., $\rgsat{\symbpes(\symbcore)}{\rgcond{pres_\symbcore}{Rs_\symbcore}{Gs_\symbcore}{psts_\symbcore}}$. 
To show the premise (2), it suffices to show that (2.1) each action transition of an event preserves the guarantee condition of the event, and (2.2) the guarantee condition of all events preserves $I$. To show the premise (2.1), it suffices to show that rely-guarantee conditions of event systems are compatible, i.e., for any $\symbcore, \symbcore'$ such that $\ \symbcore \neq \symbcore'$, $Gs_\symbcore$ is a subset of $Rs_{\symbcore'}$. Moreover, we only consider computations of $\symbpes$ with an initial state in $init$ and without environment transitions. These premises are those of the $\textsc{Par}$ proof rule by reduction of $\rgsat{\symbpes}{\rgcond{Init}{\{\}}{UNIV}{UNIV}}$. We specify the guarantee and post-conditions as $UNIV$ to automatically ensure the premises of the $\textsc{Par}$ proof rule, that is for any $\symbcore$ such that  $\ Gs_\symbcore$ is a subset of $G$ and $psts_\symbcore$ is a subset of $pst$ . 
Therefore, we have the following theorem for invariant verification:

%Then, by the compositionality of the semantics in {\lemmaprefix} \ref{lm:semaitc_compositional}, each transition of $\symbpes$ is a transition of one of its event systems. Furthermore, by {\lemmaprefix} \ref{lm:seri} each transition of an event system is a transition of its events.

%\begin{theorem}[Invariant Verification]
%\label{thm:invariant}
%For any $\symbpes$, $Init$, and $I$, if (1) $Init \subseteq I$, (2) $\rgsat{\symbpes}{\rgcond{Init}{\{\}}{UNIV}{UNIV}}$, and (3) the guarantee condition of each event in $\symbpes$ is stable for $I$, i.e., $\forall \symbevt \in evts(\symbpes). \ stable(I,guar(\Gamma(\symbevt)))$, then $I$ is an invariant of $\symbpes$ w.r.t. $Init$. 
%\end{theorem}

\begin{theorem}[Invariant Verification]
\label{thm:invariant}
For any $\symbpes$, $Init$, and $I$, if 
\begin{itemize}
\item $\rgsat{\symbpes}{\rgcond{Init}{\{\}}{UNIV}{UNIV}}$.
\item $Init \subseteq I$.
%\item Events defined in $\symbpes$ are basic events, i.e., $\forall \symbevt \in evts(\symbpes). \ is\_basic(\symbevt)$.
\item The guarantee condition of each event in $\symbpes$ is stable for $I$, i.e., 
\vspace{-6pt}
\[\forall \symbevt \in evts(\symbpes). \ stable(I,guar(\Gamma(\symbevt)))\]
\end{itemize}
\vspace{-6pt}
then $I$ is an invariant of $\symbpes$ w.r.t. $Init$. 
\end{theorem}

We give the rely-guarantee specification for each event in $\symbpes$ by a function $\Gamma$. $\Gamma(ev)$ is the rely-guarantee specification of the event $ev$, and $guar(\Gamma(ev))$ is the guarantee condition in its specification.

%Since $\anonevt{\symbprog}$ is only used to represent intermediate execution, all events in the specification of $\symbpes$ are basic events. 

%As the considered system is closed, the rely condition of $\symbpes$ is empty. For safety, we extremely relax the guarantee and post conditions to $UNIV$, which is always satisfied in the $\textsc{Par}$ proof rule and simplifies the reasoning. Since $\anonevt{\symbprog}$ is only to represent intermediate execution, all events in the specification of $\symbpes$ are basic ones. 

\section{Case Studies}
\label{sect:study_cases}
\begin{figure}[t]
\begin{center}
\includegraphics[width=4.6in]{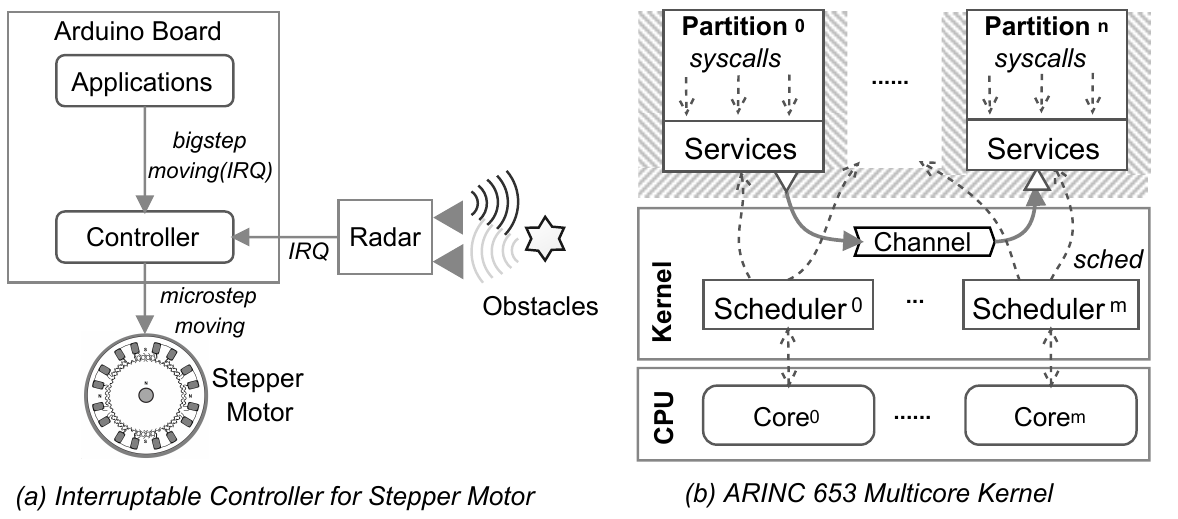}
\end{center}
\caption{Architectures of Case Studies}
\label{fig:study_cases}
\zipfigaft
\end{figure}

We develop formal specifications as well as their correctness and invariant proof in Isabelle/HOL for the two case studies. The architectures of them are shown in {\figprefix} \ref{fig:study_cases}. 

The first case study is a demo of an unmanned vehicle using a stepper motor. The application software and the controller are deployed on an Arduino development board. The controller provides bigstep system calls to applications, such as moving 10 meters forward. In the system calls, the controller drives the stepper motor in microstep mode. Since obstacles may appear at any time and be detected by a radar during movement of the vehicle, the execution of system calls in the controller has to be interruptible to avoid collision. 

ARINC 653 \cite{ARINC653p1_4} is the premier safety standard of partitioning OSs targeted at multicore processing environments and has been complied with by mainstream industrial implementations. A kernel instance executes on each physical core to manage and schedule the deployed partitions on it. System calls are invoked by programs in partitions by triggering a syscall interrupt. The scheduling is triggered by a timer interrupt. A multicore kernel is thus considered as a reactive system to these interrupts. Interrupt handlers are executed in parallel on processor cores and may access shared resources (e.g., communication channels). Note that device drivers are deployed in partitions and execute by invoking system calls of ARINC 653. 

We create the concrete syntax of {\slang} in Isabelle/HOL to ease the development of system specification. 
In the concrete syntax, the statement ``\stmtatom{c}'' denotes an atomic program, which is syntactically ``\stmtawait{\textit{True}}{c}''. ``$\event{(l,g,P)}$'' is denoted as ``$\stmtevent{l}{g}{P}$'' . When there is no ``$\cmdfont{WHEN} \ g$'' part, it means that the guard condition is $True$. 
A parameterized event is defined as ``$\lambda (elabel, plist, \symbcore). \  \event{(l,g,P)}$'', where $elabel$ is an identifid label and $plist$ is a list of parameters. $elable$, $plist$ and $\symbcore$ are thus variables in the event declaration of $\event{(l,g,P)}$. In the two case studies, we instantiate the event name $l$ to a tuple $(elabel, plist, \symbcore)$ which is syntactically represented as ``elabel [ plist ] @ $\symbcore$''. 
%
%For convenience, we first show parts of concrete syntax for in {\slang}. 
%In the {\cmdfont{Basic}} statement, we use the $\_update\_name$ function in Isabelle to update the state space. 
%The \cmdfont{ATOM} statement denotes an atomic program. 
%A parameterized event is defined as ``$\lambda (elabel, plist, \symbcore). \  \event{(l,g,P)}$'', where $elabel$ is an identifid label and $plist$ is a list of parameters. $elable$, $plist$ and $\symbcore$ are thus variables in the event declaration of $\event{(l,g,P)}$. In the two case studies, we instantiate the event name $l$ to a tuple $(elabel, plist, \symbcore)$ which is syntactically represented as ``elabel [ plist ] @ $\symbcore$''. 
%
%
%$\fontsize{8pt}{0cm}
%\begin{aligned}
%%SKIP \equiv \cmdbasic{Id} \quad \quad & \thes{x} := a \equiv \cmdbasic{\isasymguillemotleft \ \thes{\ ((\_update\_name \ x \ (\lambda\_. \ a))} \isasymguillemotright} \\
%\thes{x} := a &\equiv \cmdbasic{\isasymguillemotleft \ \thes{\ ((\_update\_name \ x \ (\lambda\_. \ a))} \isasymguillemotright} \\
%%\stmtif{b}{c1}{c2} &\equiv \cmdcond{\stset{b}}{c1}{c2} \\
%%\stmtifa{b}{c} &\equiv \stmtif{b}{c1}{SKIP} \\
%%\stmtwhile{b}{c} &\equiv \cmdwhile{\stset{b}}{c} \\
%%\stmtawait{b}{c} &\equiv \cmdawait{\stset{b}}{c} \\
%\stmtatom{c} &\equiv \stmtawait{True}{c} \\
%\stmtevent{l}{g}{P} &\equiv \event{(l,g,P)} \\
%\stmteventa{l}{P} &\equiv \stmtevent{l}{True}{P}
%\end{aligned}
%$

%\subsection{Concrete Syntax of {\slang} Language}
%\label{subsect:conc_syntax}

\subsection{Interruptible Controller for Stepper Motor}
\label{subsect:interrupt_case}

\begin{figure}[t]
%\hspace{1.0cm}
\centering
\begin{minipage}[t]{0.8\textwidth}
\begin{isabellec} \fontsize{8pt}{0cm} %\footnotesize %\scriptsize
\specrg{$\stset{UNIV}$}  \isanewline
\isacommand{EVENT} forward \ [Natural v] @ C \ \isacommand{THEN} \isanewline
\specrg{$\stset{UNIV}$}  \isanewline
\quad \stmtirq{C}{i := 0};;  \ 
\specb{\forall V. \vdash i:=0 \ \mathbf{sat} \ \langle \stset{hd \ \thes{stack} = C}\cap \{V\},Id,UNIV,\{s | (V,s) \in G\} \cap \stset{\thes{i} = 0}\rangle} 
\specrg{$\stset{\thes{i} = 0}$}  \isanewline
\quad \stmtirq{C}{pos\_aux := car\_pos};; \ \ \  
\specb{\forall V. \vdash \ pos\_aux := car\_pos \ \mathbf{sat} \ \langle \stset{hd \ \thes{stack} = C \wedge \thes{i} = 0}\cap \{V\},} \isanewline
\quad \quad \quad \quad \quad \quad \quad \quad \quad \quad \quad \quad \quad \quad \quad  
\specb{Id,UNIV,\{s \mid (V,s) \in G\} \cap \stset{\thes{car\_pos} = \thes{pos\_aux} + \thes{i}}\rangle}\isanewline
\specrg{$\stset{\thes{car\_pos} = \thes{pos\_aux} + \thes{i}}$} \isanewline
\quad \isacommand{WHILE} i $\neq$ int \ v $\wedge$ $\neg$ collide (car\_pos + 1) \ obstacle\_pos \isacommand{DO} \isanewline
\quad \quad \specb{\stset{\thes{car\_pos} = \thes{pos\_aux} + \thes{i}} \cap - \stset{i \neq int \ v \wedge \neg collide (car\_pos + 1) \ obstacle\_pos} \subseteq Q} \isanewline
\specrg{$\stset{\thes{car\_pos} = \thes{pos\_aux} + \thes{i} \wedge \thes{i} \neq int \ v \wedge \neg collide  \ (\thes{car\_pos} + 1) \ \thes{obstacle\_pos} }$}  \isanewline
\quad \quad \stmtirq{C}{\stmtatom{
%\isanewline
%\specrg{$\stset{\thes{car\_pos} = \thes{pos\_aux} + \thes{i} \wedge \thes{i} \neq int \ v \wedge \neg collide  \ (\thes{car\_pos} + 1) \ \thes{obstacle\_pos} \wedge  hd \ \thes{stack} = C}$}  \isanewline
\quad %\quad \quad \quad  \quad
\isacommand{IF} $\neg$ collide (car\_pos + 1) \ obstacle\_pos \isacommand{THEN} 
\isanewline
\specrg{$\stset{\thes{car\_pos} = \thes{pos\_aux} + \thes{i} \wedge \thes{i} \neq int \ v \wedge \neg collide  \ (\thes{car\_pos} + 1) \ \thes{obstacle\_pos} \wedge hd \ \thes{stack} = C}$}  \isanewline
\quad \quad \quad \quad \quad \quad car\_pos := car\_pos + 1 \isanewline
\quad \quad \quad \quad \quad \quad \quad
\specb{\forall V\ T. \vdash \ car\_pos := car\_pos + 1 \ \mathbf{sat} \ \langle \llbrace \thes{car\_pos} = \thes{pos\_aux} + \thes{i} \wedge \thes{i} \neq int \ v} \isanewline
\quad \quad \quad \quad \quad \quad \quad \quad \quad
\specb{ \wedge \neg collide  \ (\thes{car\_pos} + 1) \ \thes{obstacle\_pos} \wedge hd \ \thes{stack} = C \rrbrace \cap \{V\} \cap \{T\},}\isanewline
\quad \quad \quad \quad \quad \quad \quad \quad \quad \quad \quad \quad
\specb{Id,UNIV,\{s \mid (V,s) \in G \} \cap \stset{\thes{car\_pos} = \thes{pos\_aux} + \thes{i} + 1} \rangle }\isanewline
\specrg{$\stset{\thes{car\_pos} = \thes{pos\_aux} + \thes{i} + 1}$}  \isanewline
\quad \quad \quad \quad \quad \isacommand{FI} %\isanewline \quad \quad \quad
}};; \isanewline
\specrg{$\stset{\thes{car\_pos} = \thes{pos\_aux} + \thes{i} + 1}$}  \isanewline
\quad \quad \stmtirq{C}{i := i + 1} \ \ \specb{\forall V. \vdash i := i + 1 \ \mathbf{sat} \langle \stset{\thes{car\_pos} = \thes{pos\_aux} + \thes{i} + 1 \wedge	hd \ \thes{stack} = C}\cap \{V\},} \isanewline
\quad \quad \quad \quad \quad \quad \quad \quad \quad \quad \quad \quad \quad \quad \quad  
\specb{Id,UNIV,\{s \mid (V,s) \in G\} \cap \stset{\thes{car\_pos} = \thes{pos\_aux} + \thes{i}} \rangle} \isanewline
\specrg{$\stset{\thes{car\_pos} = \thes{pos\_aux} + \thes{i}}$}  \isanewline
\quad \isacommand{OD};; \isanewline
\specrg{$\stset{\thes{car\_pos} = \thes{pos\_aux} + \thes{i} \wedge (\thes{i} = int \ v \vee collide \ (\thes{car\_pos} + 1) \ obstacle\_pos)}$}  \isanewline
\quad \stmtirq{C}{iret} \quad  \specb{\forall V. \rgsat{iret}{\rgcond{Q \cap \stset{hd \ \thes{stack} = C}\cap \{V\}}{Id}{UNIV}{\{s \mid (V,s) \in G \} \cap Q}}} \isanewline
\specrg{$\stset{\thes{car\_pos} = \thes{pos\_aux} + \thes{i} \wedge (\thes{i} = int \ v \vee collide \ (\thes{pos\_aux} + \thes{i} + 1) \ obstacle\_pos)}$}  \isanewline
\isacommand{END} \isanewline
\specrg{$\stset{\thes{car\_pos} = \thes{pos\_aux} + \thes{i} \wedge (\thes{i} = int \ v \vee collide \ (\thes{pos\_aux} + \thes{i} + 1) \ obstacle\_pos)}$}
\end{isabellec}
\end{minipage}

%\caption{Proof Sketch of $\ \ \rgsat{forward \ v}{forward\_RGCondition \ v}$ }
\caption{Definition of Event $forward$ and its Proof Sketch}
\label{fig:forwardproof}
\zipfigaft
%\vspace{-2mm}
\end{figure}

In the case study, we apply our approach to preemption and multi-level interrupts. We prove functional correctness and safety properties of an unmanned vehicle with a controller for a stepper motor and a radar to detect obstacles. The architecture of the case is illustrated in {\figprefix} \ref{fig:study_cases} (a). 
We assume that the vehicle can move forward or backward. Thus, we consider two system calls of the controller, i.e. \textit{forward(nat v)} and \textit{backward(nat v)}, where \textit{v} is the distance to move. When the controller received a system call, it drives the stepper motor in the microstep mode until moving over the distance. When the vehicle is moving, obstacles may appear and be detected by the radar. To avoid collision, the radar then sends an IRQ to the controller to interrupt the movement. 

We design three concurrent modules: a radar (R), a controller (C), and a programmable interrupt controller (PIC). The two system calls and the reaction to radar are interrupt handlers. System calls from applications and detections from radar will send IRQs to the PIC. Then the PIC blocks the current handler and jumps to a new one. In order to represent the multi-level interrupts, we define a stack to save the IRQs and use a guarded statement $\stmtirqa{C}{p} \equiv \stmtawait{hd \ stack = C}{p}$ to represent that an internel step $p$ of a handler $C$ can be executed only when the handler is the top element of the stack, i.e. the currently executing handler. 
The system state is defined by a  \textit{car\_pos} variable showing the current position of the vehicle; an \textit{obstacle\_pos} saving the positions of all obstacles detected; and auxiliary variables \textit{i}, \textit{pos\_aux}, and \textit{obst\_pos\_aux} locally used in the events.

%\begin{isabellec}\centering
%\vspace{6pt}
%\stmtirq{C}{p} $\equiv$ \stmtawait{hd stack = C}{p}
%\vspace{6pt}
%\end{isabellec}

We define a set of events to specify the interrupt handlers for system calls, detected obstacle, and IRQ to PIC. The event $forward$ is the handler of the system call ``forward(nat distance)'' shown in {\figprefix} \ref{fig:forwardproof} in black color. 
%The statement ``\stmtatom{...}'' denotes an atomic program, which is syntactically ``\stmtawait{\textit{True}}{...}''. 
The event $backward$ is the handler of the system call ``backward(nat distance)'', which is similar to $forward$ and not presented here. The two handlers drive the stepper motor to move one step forward or backward each time until it moves over $v$ steps. During the movement, if the handler finds that there is an obstacle in the next position, it stops immediately. In the end of the handler, the \textit{iret} statement pops the IRQ stack. 
The event $obstacle$ shown as follows is the handler of IRQs from the radar. The event will insert the position of the appeared obstacle into \textit{obstacle\_pos}. Here, we assume that an obstacle will not appear at the current position of the vehicle as well as one step before and after the current position. Otherwise, the vehicle cannot avoid collision in time. 
The event $IRQs$ simulates the receiving of IRQs in the PIC. It pushes the new IRQ to the stack. We assume that if an IRQ from one device is being handled, no IRQ from the same device will come. The two events are shown as follows. 
In this case, we permit that events $forward$ (or $backward$) and $obstacle$ could be preempted by each other. %We assure the correctness and safety of them. 

\vspace{2mm}
\begin{minipage}[t]{1.0\textwidth}\hspace{-0.5cm}
\begin{minipage}[t]{0.60\textwidth}
\begin{isabellec} \fontsize{7pt}{0cm} %\footnotesize %\scriptsize
\isacommand{EVENT} obstacle \ [Integer v] @ R \isanewline
\isacommand{THEN} \isanewline
\quad \stmtirq{R}{obst\_pos\_aux :=  obstacle\_pos};; \isanewline
\quad \stmtirq{R}{
\isacommand{IF} \ v $\neq$ car\_pos $\wedge$ v $\neq$ car\_pos + 1 %\isanewline
$\wedge$ v $\neq$ car\_pos - 1 \ \isacommand{THEN} \isanewline
\quad \quad \quad \quad obstacle\_pos :=  v \# obstacle\_pos \isanewline
\quad \quad \quad  \isacommand{FI};; \isanewline
}
\quad \stmtirq{R}{iret} \isanewline
\isacommand{END}
\end{isabellec}
\end{minipage}
~
\begin{minipage}[t]{0.38\textwidth}
\begin{isabellec} \fontsize{8pt}{0cm} %\footnotesize %\scriptsize
\isacommand{EVENT} IRQs \ [Irq d] @ PIC \isanewline
\isacommand{THEN} \isanewline
\quad \stmtatom{
%\isanewline \quad \quad 
\ \ 
\isacommand{IF} hd \ stack $\neq$ d \ \isacommand{THEN} \isanewline
\quad \quad \quad push \ d \isanewline
\quad \quad \isacommand{FI} \isanewline \quad 
} \isanewline
\isacommand{END}
\end{isabellec}
\end{minipage}
\end{minipage}
\vspace{2mm}

The system is the parallel composition of the controller, the radar, and the PIC, shown as follows. Since the events can be triggered with any parameter, we use the union of events w.r.t. the parameters. 

\vspace{2mm}
\begin{minipage}[t]{1.0\textwidth}
\hspace{-10mm}
\centering \fontsize{9pt}{0cm}
$\begin{aligned}
& Ctrl \equiv ({\bigcup}_{v} forward \ v) \cup ({\bigcup}_{v} backward \ v) \quad \quad Radar \equiv ({\bigcup}_{v} obstacle \ v) \quad \quad PIC \equiv ({\bigcup}_{r} IRQs \ r)
\end{aligned}$
\end{minipage}
\vspace{-1mm}

The functional correctness of the system is specified and verified by concerning the rely-guarantee conditions of their events. We define the rely-guarantee conditions of each event. As an example, the condition of $forward$ is defined as follows. 
The expression $\stset{\phi}$ is a concrete syntax for a set of states satisfying $\phi$. We present the value of a variable $x$ in the state by $\thes{x}$. We present the value of a variable $x$ in the state before and after a transition by $\bfs{x}$ and $\afs{x}$ respectively. 
The pre-condition is relaxed to the universal set $\stset{True}$. 
The rely condition shows that $car\_pos$ and the two local variables ($i$ and $pos\_aux$) are not changed by the environment. 
Moreover, during handling radar IRQs ($hd \ \bfs{stack} \neq C$), the rely condition includes state changes in the events $obstacle$ and $IRQs$. In first case, $obstacle\_pos$ remains unchanged during stack operation ($\afs{stack} = tl \bfs{stack} \vee \afs{stack} = C \# \bfs{stack}$) and assignment of local variables ($\afs{obst\_pos\_aux} = \bfs{obstacle\_pos}$). In second case, new obstacles may occur and thus $set \ \bfs{obstacle\_pos} \subseteq set \ \afs{obstacle\_pos}$. As we assume that obstacles will not appear at one step forward the vehicle, the rely condition also requires that collision at $\afs{car\_pos} + 1$ is the same for before and after obstacles occurring.  
If no new obstacles are detected, i.e., the controller is currently running ($hd \ \bfs{stack} = C$), the environment does not change the variables $obstacle\_pos$ and $obst\_pos\_aux$. The PIC may receive new IRQs from the Radar ($\afs{stack} = R \# \bfs{stack}$), and then $forward$ is interrupted.
The post-condition defines the correctness of the $forward$ event. It means that the vehicle will be moved $i$ steps forward. If no obstacle appears among the distance $v$, then $i = v$. Otherwise, the vehicle stops before the obstacle, i.e. $collide \ (\thes{pos\_aux} + \thes{i} + 1) \ obstacle\_pos$.

%\begin{figure*}
%\vspace{-5mm}
%\fontsize{8pt}{0cm}
\vspace{2mm}

\begin{minipage}[t]{1.0\textwidth} \fontsize{8.5pt}{0cm}
\hspace{-6mm}
\centering
$\begin{aligned} 
&forward\_RGCondition \ v \equiv 
\langle \stset{ True },  \llbrace \afs{car\_pos} = \bfs{car\_pos} \wedge \afs{i} = \bfs{i} \wedge \afs{pos\_aux} = \bfs{pos\_aux}  \\
& \wedge (hd \ \bfs{stack} \neq C \longrightarrow (\afs{obstacle\_pos} = \bfs{obstacle\_pos} \wedge (\afs{stack} = tl \bfs{stack} \vee \afs{stack} = C \# \bfs{stack} \\
& \quad \quad \quad \quad \quad \quad \vee \afs{obst\_pos\_aux} = \bfs{obstacle\_pos})) \vee (set \ \bfs{obstacle\_pos} \subseteq set \ \afs{obstacle\_pos} \wedge \\
& \quad \quad \quad \quad \quad \quad collide \ (\afs{car\_pos} + 1) \ \bfs{obstacle\_pos} = collide \ (\afs{car\_pos} + 1) \ \afs{obstacle\_pos})) \\
& \wedge (hd \ \bfs{stack} = C \longrightarrow \afs{obstacle\_pos} = \bfs{obstacle\_pos} \wedge \afs{stack} = R \# \bfs{stack} \wedge \bfs{obst\_pos\_aux} = \afs{obst\_pos\_aux})\rrbrace \cup Id , \\
& \llbrace (((\afs{i} = 0 \vee \afs{i} = \bfs{i} + 1 \vee \afs{stack} = tl \ \bfs{stack}) \wedge \afs{car\_pos} = \bfs{car\_pos}) \vee 
 (\neg collide  \ (\bfs{car\_pos} + 1) \ \bfs{obstacle\_pos} \\
& \wedge \afs{car\_pos} = \bfs{car\_pos} + 1)) \wedge hd \ \bfs{stack} = C  
 \wedge \afs{obstacle\_pos} = \bfs{obstacle\_pos} \wedge \bfs{obst\_pos\_aux} = \afs{obst\_pos\_aux} \rrbrace \cup Id, \\
& \llbrace \thes{car\_pos} = \thes{pos\_aux} + \thes{i} \wedge (\thes{i} = \ v \vee collide \ (\thes{pos\_aux} + \thes{i} + 1) \ obstacle\_pos) \rrbrace \rangle
\end{aligned}$
%\vspace{-16pt}
%\end{figure*}
\end{minipage}
\vspace{2mm}

Functional correctness of the $forward$ event is proven by induction of rely-guarantee proof rules. We use $R$ and $G$ to denote the rely and guarantee conditions of $forward\_RGCondition$ respectively. {\figprefix} \ref{fig:forwardproof} shows the proof sketch of $forward$ satisfies its rely-guarantee condition. In the figure, pre- and post-conditions of each statement are shown in blue colour, and verification conditions produced  by proof rules in green colour.  We omit general proof obligations in rely-guarantee, i.e., stability of pre- and post-conditions w.r.t. the rely relation, and that rely and guarantee are reflexive. We use a loop invariant $\stset{\thes{car\_pos} = \thes{pos\_aux} + \thes{i}}$ for the \textbf{WHILE} statement.
Then, the functional correctness of \textit{Ctrl}, \textit{Radar}, and \textit{PIC} could be proved using the $\textsc{EvtSet}$ proof rule in {\figprefix} \ref{fig:proofrule}. Finally, we define the rely-guarantee condition of the whole system as $\rgcond{UNIV}{\{\}}{G_{sys}}{Q_{sys}}$ and prove the correctness using the $\textsc{Par}$ proof rule. % where $G_{sys}$ and $Q_{sys}$ are calculated according to the $\textsc{Par}$ proof rule. 
We release the pre-condition to the universal set and consider the system as a closed one, i.e. the rely condition is an empty set.

We verify a safety property defined as $inv \equiv \stset{\neg collide \ \ \thes{car\_pos}\ \ \ \thes{obstacle\_pos})}$ of the system, which means that the vehicle will not collide with any obstacle at any time. By the functional correctness of the system and the $\textsc{Conseq}$ proof rule, it is straightforward that $\rgsat{VehicleSpec}{\rgcond{\{s_0\}}{\{ \}}{UNIV}{UNIV}}$, where $s_0$ is the initial state of the system. Then we prove that $inv$ is an invariant of $VehicleSpec$ using {\theoremprefix} \ref{thm:invariant} by showing that $\{s_0\} \subseteq inv$ and the guarantee condition of each event is stable for $inv$.  

%\vspace{-8pt}
%\[
%inv \equiv \stset{\neg collide \ \ \thes{car\_pos}\ \ \ \thes{obstacle\_pos})}
%\]

In the case study, we have only one external device, i.e. the radar. It is straightforward to support multiple devices which could be specified as event systems in {\slang}.

\subsection{IPC in ARINC 653 Multicore OS Kernels}
\label{subsect:ipc_case}
%\qmark{emphasize the parametrization for any core, any parameters}

This case study concerns multicore concurrency and invariant verification. Since device drivers run in special partitions, we do not consider multi-level interrupts in the kernel. 
As shown in {\figprefix} \ref{fig:study_cases} (b), IPC in ARINC 653 is conducted via messages on channels configured among partitions. Partitions have access to channels via \emph{ports} which are the endpoints of channels. A queuing mode channel has a source port, a destination port, and a bounded FIFO message queue. 

The kernel configuration is split into static and dynamic components in Isabelle. We create a constant $conf$ to be used in the specification defining static components of the state. In $conf$, $c2s$ is the mapping from cores to schedulers and is bijective. $p2s$ is the deployment of partitions to schedulers. $p2p$ indicates the partition a port belongs to. $chsrc$ and $chdest$ indicate the source port and destination port of a queuing channel. Finally, $chmax$ defines the maximum capacity of a channel.
Our specification is created based on abstract data types for these elements: $Core$, $Part$, $QChannel$, $Port$, $Message$. It means that we conduct formal verification on arbitrary system configurations rather than a concrete instance. 
 
The dynamic component of the kernel state concerns states of schedulers, channels, and partitions. The state of a scheduler shows the current partition under execution. The state of a channel keeps the information of the messages in the FIFO queue. The state of a partition is defined as IDLE, READY or RUN. We define $s_0$ as the initial state of the system.

\vspace{2mm}
\begin{isabellec} \fontsize{9pt}{0cm}
%\isacodeftsz
\ \ \ \ \isacommand{record}\isamarkupfalse%
\ Config {\isacharequal}\ c2s\ {\isacharcolon}{\isacharcolon}\ Core\ {\isasymRightarrow} \ Sched
\ \ \ \ p2s\ {\isacharcolon}{\isacharcolon}\ Part\ {\isasymRightarrow} \ Sched \ \ \ \
p2p \ {\isacharcolon}{\isacharcolon}\ Port \ {\isasymRightarrow} \ Part \isanewline
\ \ \ \ \ \ \ \ \ \ \ \ \ \ \ \ \ \ \ \ \ \ \ \ \ \ \ \ \ \ \ \
chsrc \ {\isacharcolon}{\isacharcolon}\ QChannel \ {\isasymRightarrow} \ Port \ \ \ \ 
chdest \ {\isacharcolon}{\isacharcolon}\ QChannel \ {\isasymRightarrow} \ Port \isanewline
\ \ \ \ \ \ \ \ \ \ \ \ \ \ \ \ \ \ \ \ \ \ \ \ \ \ \ \ \ \ \ \
chmax  \ {\isacharcolon}{\isacharcolon}\ QChannel \ {\isasymRightarrow} \ nat

\ \ \ \ \isacommand{record}\isamarkupfalse%
\ State {\isacharequal}\ cur \ {\isacharcolon}{\isacharcolon}\ Sched {\isasymRightarrow} \ Part \ option \ \ \ \ 
qbuf \ {\isacharcolon}{\isacharcolon}\ QChannel \ {\isasymRightarrow} \ Message list \isanewline
\ \ \ \ \ \ \ \ \ \ \ \ \ \ \ \ \ \ \ \ \ \ \ \ \ \ \ \ \ \ 
qbufsize \ {\isacharcolon}{\isacharcolon}\ QChannel \ {\isasymRightarrow} \ nat \ \ \ \ 
partst \ {\isacharcolon}{\isacharcolon}\ Part \ {\isasymRightarrow} \ PartMode              
%\ \ \ \ \ \ \ \ \ \ \ \ \ \ \ \ \ \ \ \ \ \ \ \ \ \ \ \ \ \ 
%slock \ {\isacharcolon}{\isacharcolon}\ SampChannel \ {\isasymRightarrow} \ bool
\end{isabellec}
\vspace{2mm}

We define a set of events to specify the scheduling and communication services. These events are parametrized by their input parameters and the core identifier $\symbcore$. The $Schedule$ and $Send\_QMsg$ events are shown below. A partition $p$ can be scheduled on processor core $\symbcore$ when $p$ was deployed on $\symbcore$ and the state of $p$ is not IDLE. The event first sets the state of the currently running partition to READY and current partition on $\symbcore$ to $None$. Then it sets $p$ as the current partition and its state to RUN. 
The $Send\_QMsg$ event can happen in the current partition on processor core $\symbcore$ when the source port $p$ is configured in the current partition. The event will be blocked until the message queue of the operated channel has available spaces. Then it inserts the message into the tail of the message queue and increases the size of the queue. 

\vspace{2mm}
\begin{minipage}[t]{1.0\textwidth} \hspace{-3mm}
\begin{minipage}[t]{0.5\textwidth}
\begin{isabellec} \fontsize{8pt}{0cm} %\footnotesize %\scriptsize
\isacommand{EVENT} Schedule \ [Part p] @ $\symbcore$ \isanewline
\isacommand{WHEN}  p2s conf p = c2s conf $\symbcore$ $\wedge$ partst p $\neq$ IDLE \isanewline
%\ \ \ \ \ \ $\wedge$ (cur((c2s conf) $\symbcore$) = None $\vee$ p2s conf (the (cur((c2s conf) $\symbcore$))) = c2s conf $\symbcore$) \isanewline
\isacommand{THEN} \isanewline
\quad \isacommand{IF} cur((c2s conf) $\symbcore$) $\neq$ None \isacommand{THEN} \isanewline
\quad \quad \stmtatom{\isanewline \quad \quad \quad partst := partst(cur ((c2s conf) $\symbcore$) := READY);; \isanewline 
\quad \quad \quad cur := cur((c2s conf) $\symbcore$ := None)  \isanewline \quad \ } \isanewline
\quad \isacommand{FI};; \isanewline
\quad \stmtatom{\isanewline \quad \quad cur := cur((c2s conf) $\symbcore$ := Some p);; \isanewline \quad \quad partst := partst(p := RUN) \isanewline \ \ } \isanewline
\isacommand{END}
\end{isabellec}
\end{minipage}
~
\begin{minipage}[t]{0.52\textwidth}
\begin{isabellec} \fontsize{8pt}{0cm} %\footnotesize %\scriptsize
\isacommand{EVENT} Send\_QMsg \ [Port p, Msg m] @ $\symbcore$ \isanewline
\isacommand{WHEN}  is\_src\_port conf p $\wedge$ cur (c2s conf $\symbcore$) $\neq$ None \isanewline
\quad \quad \quad \quad $\wedge$ port\_of\_part conf p (cur (c2s conf $\symbcore$)) \isanewline
\isacommand{THEN} \isanewline
\quad \isacommand{AWAIT} qbufsize (ch\_sport conf p) \isanewline
\quad \quad \quad \quad \quad $<$ chmax conf (ch\_sport conf p) \isanewline
\quad \isacommand{THEN} \isanewline
\quad \quad qbuf := qbuf (ch\_sport conf p \isanewline
\quad \quad \quad \quad \quad \quad := qbuf (ch\_sport conf p) @ [m]);; \isanewline
\quad \quad qbufsize := qbufsize (ch\_sport conf p \isanewline
\quad \quad \quad \quad \quad \quad:= qbufsize (ch\_sport conf p) + 1) \isanewline
\quad \isacommand{END} \isanewline
\isacommand{END}
\end{isabellec}
\end{minipage}
\end{minipage}
\vspace{2mm}

The parallel event system in {\slang} of multicore kernels is thus defined as follows. Each core deploys the same event sequence parametrized with a core identifier $\symbcore$. When starting a kernel instance on each processor core, we use the event $Core\_Init$ to initialize the kernel state of each core. Then, the kernel instance reacts to system calls as defined in $Esys \ \symbcore$, which is an event set. 
%The event systems defined on each core are the same. 

%\vspace{6pt}
%\begin{tabular}{l} \fontsize{8pt}{0cm}
%$ARINCSpec \equiv \lambda \symbcore. \ (\evtseq{(Core\_Init \ \symbcore)}{(Esys \ \symbcore)})$
%\\
%\fontsize{8pt}{0cm}$
%\begin{aligned}
%Esys \ \symbcore \equiv \lambda \symbcore. \ ({\bigcup}_{p} Schedule \ \symbcore\ p) & \cup ({\bigcup}_{(p,m)} Send\_Que\_Msg \ \symbcore \ p \ m) \\
%& \cup ({\bigcup}_{p} Recv\_Que\_Msg \ \symbcore \ p)
%\end{aligned}
%$
%\end{tabular}

%\begin{figure*}
%\vspace{-18pt}
\vspace{5mm}
\begin{minipage}[t]{1.0\textwidth}
\hspace{-10mm}
\centering \fontsize{9pt}{0cm}
$
\begin{aligned}
& ARINCSpec \equiv \lambda \symbcore. \ (\evtseq{(Core\_Init \ \symbcore)}{(Esys \ \symbcore)}) \\
& Esys \ \symbcore \equiv \lambda \symbcore. \ ({\bigcup}_{p} Schedule \ \symbcore\ p) \cup ({\bigcup}_{(p,m)} Send\_QMsg \ \symbcore \ p \ m)  \cup ({\bigcup}_{p} Recv\_Que\_Msg \ \symbcore \ p)
\end{aligned}
$
%\vspace{-18pt}
%\end{figure*}
\end{minipage}
\vspace{1mm}

For the purpose of compositional reasoning based on events, we specify the rely-guarantee conditions of each event. The conditions of $Send\_QMsg$ are defined as follows. 
The event executing on $\symbcore$ relies on the current partition on $\symbcore$ not being changed by events on other cores. The guarantee condition shows that internal steps of the event will not change the current partition on any core, the state of any partition, as well as the message queue and size of other channels. The event also guarantees that if the size is equal to the number of messages before an internal step, then they are still equal after the step. 

%\begin{figure*}
%\vspace{-16pt}
\vspace{1mm}
\begin{minipage}[t]{1.0\textwidth}
\hspace{-10mm}
\fontsize{9pt}{0cm}
\centering
$\begin{aligned} 
Send&\_QMsg\_RGCondition \equiv
\langle \stset{True}, \stset{ \afs{cur} \ (c2s \ conf \ k) = \bfs{cur} \ (c2s \ conf \ k) }, \\
& \llbrace \afs{cur} = \bfs{cur} \wedge \afs{partst} = \bfs{partst} \wedge \\
& \quad (\bfs{qbufsize} \ (ch\_srcqport \ conf \ p) = length \ (\bfs{qbuf} \ (ch\_srcqport \ conf \ p)) \\
& \quad \quad \quad \longrightarrow \afs{qbufsize} \ (ch\_srcqport \ conf \ p) = length \ (\afs{qbuf} \ (ch\_srcqport \ conf \ p))  ) \\
& \wedge (\forall \ c. \ c \neq ch\_srcqport \ conf \ p \longrightarrow \afs{qbuf} \ c = \bfs{qbuf} \ c) \\
& \wedge (\forall \ c. \ c \neq ch\_srcqport \ conf \ p \longrightarrow \afs{qbufsize} \ c = \bfs{qbufsize} \ c) \rrbrace, \stset{ True } \rangle
\end{aligned}$
%\vspace{-16pt}
%\end{figure*}
\end{minipage}
\vspace{1mm}

We carry out verification of a system invariant $inv$. The invariant \textit{inv1} means that if a partition $p$ is the currently executing partition of a scheduler $sched$, $p$ should be deployed on the scheduler. The second one means that if a partition $p$ is deployed on a scheduler $sched$ and $p$ is the current partition, the state of $p$ is RUN. The third one defines that for any queuing channel $c$, the current size should be the number of messages in the queue.

%\begin{figure*}
%\vspace{-16pt}
\vspace{1mm}
\begin{minipage}[t]{1.0\textwidth}
\hspace{-10mm}
\fontsize{9pt}{0cm}
\centering
$\begin{aligned} 
& inv1 \equiv \stset{\forall sched \ p. \ \thes{cur} \ sched = Some \ p \longrightarrow sched = (p2s \ conf) \ p} \\
& inv2 \equiv \llbrace \forall sched \ p. \ p2s \ conf \ p = sched \wedge \thes{cur} \ sched = Some  \ p  \longrightarrow \thes{partst} \ p = RUN \rrbrace \\
& inv3 \equiv \stset{\forall c . \ \thes{qbufsize} \ c = length \ (\thes{qbuf} \ c)} \quad \quad inv \equiv inv1 \cap inv2 \cap inv3
\end{aligned}$
\end{minipage}
\vspace{1mm}
%\vspace{-16pt}
%\end{figure*}

We have that $\{s_0\} \subseteq inv$ and that all events specified in $ARINCSpec$ are basic events. Moreover, we prove that the guarantee condition of each event in $ARINCSpec$ is stable with $inv$. We have that $\rgsat{ARINCSpec}{\rgcond{\{s_0\}}{\{\}}{UNIV}{UNIV}}$ using these results and by direct application of event compositionality and events proof rules. According to {\theoremprefix} \ref{thm:invariant}, we show that $inv$ is an invariant of $ARINCSpec$ w.r.t $\{s_0\}$.

\section{Discussion and Conclusion}
\label{sect:concl}
%\qmark{``My main critism is the lack of discussion of the formalisation in Isabelle/HOL e.g. discussion of the overhead and effort involved in this formalisation. ''}

We use Isabelle/HOL as the specification and verification system for our work. The proofs are conducted in the structured proof language \emph{Isar} in Isabelle. All derivations of our proofs have passed through the Isabelle proof kernel. The statistics for the effort and size are shown in {\tableprefix} \ref{tbl:stat}, totally $\approx$ 15,000 lines of specification and proofs by a total effort of roughly 10 person-months (PM). The part of ``Language'' in the table includes definitions and their lemmas of language, semantics, and computation. For invariant verification, we prove a set of lemmas, which are also reusable for further proofs (e.g. security).

%\begin{table}[t]
%%\vspace{-8pt}
%\centering
%\footnotesize
%\caption{Specification and Proof Statistics} %\tiny,\scriptsize,\footnotesize,\small,
%\begin{tabular} {|c|c|c|c|c|c|c|c|}
%\hline
%\textbf{Item} & \textbf{Spec.} & \textbf{Proof} & \textbf{PM} & \textbf{Item} & \textbf{Spec.} & \textbf{Proof} & \textbf{PM} \\
%\hline
%\textbf{Language} & 420 & 3,000 & \multirow{3}{*}{8} & \textbf{Controller} & 260 & 900 & \multirow{2}{*}{2} \\
%\cline{1-3} \cline{5-7}
%\textbf{Proof system} & 250 & 7,100 & & \textbf{ARINC 653} & 250 & 700 & \\
%\cline{1-3} \cline{5-8}
%\textbf{Aux. Lemma/Invariant} & - & 3,400 & & \textbf{Total} & 1,180 & 15,100 & 10 \\
%\hline
%\end{tabular}
%%\vspace{-6pt}
%\label{tbl:stat}
%\end{table} 
\begin{table}[t]
%\vspace{-8pt}
\centering
\footnotesize
\caption{Specification and Proof Statistics} %\tiny,\scriptsize,\footnotesize,\small,
\begin{tabular} {|c|c|c|c|c|c|c|c|}
\hline
\textbf{Item} & \textbf{Spec.} & \textbf{Proof} & \textbf{Item} & \textbf{Spec.} & \textbf{Proof} \\
\hline
\textbf{Language} & 420 & 3,000 & \textbf{Controller} & 260 & 900 \\
\hline
\textbf{Proof system} & 250 & 7,100 & \textbf{ARINC 653} & 250 & 700 \\
\hline
\textbf{Aux. Lemma/Invariant} & - & 3,400 & \textbf{Total} & 1,180 & 15,100\\
\hline
\end{tabular}
%\vspace{-6pt}
\label{tbl:stat}
\end{table}

%In this paper, we proposed an event-based rely-guarantee approach for concurrent reactive systems. We developed two real cases of such systems, an interruptible controller for stepper motors and an ARINC 653 multicore kernels. Our approach can deal with multicore and interruptible concurrency. The approach has been mechanized in the Isabelle/HOL theorem prover and is reusable for further development. 
The Isabelle/HOL implementation of {\slang} provides a flexible and extensible framework for concurrent reactive system. As a next step of our work, we are using {\slang} and its rely-guarantee proof system to formally verify a concurrent buddy memory allocator of Zephyr RTOS at low-level design. 

In future, as an important problem in concurrent systems, formal verification of deadlock freedom in {\slang} deserves further study. Then, compositional reasoning of noninterference could be done in {\slang}. For formal development of concurrent reactive systems, event refinement is very necessary for step-wise refinement.

%%%%%%%%%%%%%%%%%%%%%%%%%%%%%%%%%%%%%%%%%%%%%%%%%%%%%%%
%%% Acknowledgements. 致谢
%%%%%%%%%%%%%%%%%%%%%%%%%%%%%%%%%%%%%%%%%%%%%%%%%%%%%%%
\Acknowledgements{We would like to thank Jean-Raymond Abrial for his suggestions. This work is supported by National Natural Science Foundation of China (NSFC) under Grant No. 61872016 and National Research Foundation, Prime Minister's Office, Singapore under its National Cybersecurity R\&D Program (Grant No. NRF2014NCR-NCR001-30).}

%%%%%%%%%%%%%%%%%%%%%%%%%%%%%%%%%%%%%%%%%%%%%%%%%%%%%%%
%%% Supplements. 补充材料, 非必选
%%%%%%%%%%%%%%%%%%%%%%%%%%%%%%%%%%%%%%%%%%%%%%%%%%%%%%%
%\Supplements{Appendix A.}

%%%%%%%%%%%%%%%%%%%%%%%%%%%%%%%%%%%%%%%%%%%%%%%%%%%%%%%
%%% Reference section. 参考文献
%%% citation in the content using "some words~\cite{1,2}".
%%% ~ is needed to make the reference number is on the same line with the word before it.
%%%%%%%%%%%%%%%%%%%%%%%%%%%%%%%%%%%%%%%%%%%%%%%%%%%%%%%
%\begin{thebibliography}{99}
%
%\bibitem{1} Author A, Author B, Author C. Reference title. Journal, Year, Vol: Number or pages
%
%\bibitem{2} Author A, Author B, Author C, et al. Reference title. In: Proceedings of Conference, Place, Year. Number or pages
%
%\end{thebibliography}

%\bibliographystyle{plain}
%\bibliography{paperbibtex}

%%%%%%%%%%%%%%%%%%%%%%%%%%%%%%%%%%%%%%%%%%%%%%%%%%%%%%%
%%% Appendix sections. 附录章节, 非必选
%%%%%%%%%%%%%%%%%%%%%%%%%%%%%%%%%%%%%%%%%%%%%%%%%%%%%%%
%\begin{appendix}
%
%
%\section{Semantics and Rely-guarantee Proof Rules}
%\label{appdx:picore_lang_sem_prfrule}
%
%\end{appendix}

\end{document}